\documentclass[twoside]{article}

\usepackage{lipsum} 

\usepackage[sc]{mathpazo} 
\usepackage[T1]{fontenc} 
\usepackage{microtype} 

\usepackage[hmarginratio=1:1,top=32mm,columnsep=20pt]{geometry} 
\usepackage{multicol} 
\usepackage[hang, small,labelfont=bf,up,textfont=it,up]{caption} 
\usepackage{booktabs} 
\usepackage{float} 
\usepackage{hyperref} 
\hypersetup{%
    pdfborder = {0 0 0}
}
\usepackage[section]{placeins}

\usepackage{lettrine} 
\usepackage{paralist} 

\usepackage{subcaption}

\usepackage{setspace}
\usepackage{amssymb,latexsym}
\usepackage{amsmath}
\usepackage{ifpdf}
\usepackage[pdftex]{graphicx}
\usepackage{amstext}
\usepackage{amssymb}
\usepackage{tocloft}
\usepackage{textcomp}

\usepackage{enumerate}
\usepackage{mathtools}
\usepackage{units}

\usepackage{bm}
\usepackage{amsbsy}
\usepackage{chapterbib}
\usepackage{cite} 
\usepackage{authblk}
\usepackage[ruled,vlined]{algorithm2e}

\usepackage{abstract} 

\usepackage{titlesec} 
\renewcommand\thesection{\arabic{section}} 
\renewcommand\thesubsection{\arabic{subsection}} 
\titleformat{\section}[block]{\large\scshape\centering}{\thesection.}{1em}{} 
\titleformat{\subsection}[block]{\large}{\thesubsection.}{1em}{} 

\usepackage{fancyhdr} 
\title{\vspace{-15mm}\fontsize{24pt}{10pt}\selectfont\textbf{Dynamic Bayesian Nonlinear Calibration}} 

\author[]{Derick L. Rivers\thanks{Corresponding author. Email:~riversdl@vcu.edu}~}
\author[]{Edward L. Boone}
\affil[]{Department of Statistical Sciences and Operations Research, Virginia Commonwealth University}
\date{}

\begin{document}

\maketitle 

\thispagestyle{fancy} 


\begin{abstract}
Statistical calibration where the curve is nonlinear is important in many areas, such as analytical chemistry and radiometry. Especially in radiometry, instrument characteristics change over time, thus calibration is a process that must be conducted as long as the instrument is in use. We propose a dynamic Bayesian method to perform calibration in the presence of a curvilinear relationship between the reference measurements and the response variable. The dynamic calibration approach adequately derives time dependent calibration distributions in the presence of drifting regression parameters. The method is applied to simulated spectroscopy data based on work by Lundberg and de Mar\'{e} (1980).

\end{abstract}



\section{Introduction}

In many areas such as analytical chemistry, bioassay, spectroscopy, and radiometry fitting a curve through data to perform statistical calibration is of great importance. The statistical calibration problem is typically carried out in two stages; first samples are collected consisting of observations and known reference measurements of a targeted subject, and second a fitted curve is established from the first stage and an observed value $y_{0}$ is used to predict an unknown targeted reference measurement $x_{0}$. The linear approach to this problem has been given much consideration from both the frequentist perspective (Eisenhart 1939; Krutchkoff 1967; Berkson 1969; Williams 1969; Halperin 1970; Martinelle 1970; Lwin and Maritz 1982) and the Bayesian perspective (Hoadley 1970; Hunter and Lamboy 1981; Eno 1999). The multivariate case to the linear calibration problem is considered by Brown (1982) from both persepectives. Bayesian dynamic approaches to the linear statistical calibration problem have been explored that consider calibration estimates as a function of time (Smith and Corbett 1987; Rivers and Boone 2014).\\
\indent Unfortunately in many cases these curves are curvilinear, and straight-line linear methods are inappropriate. Several authors have considered Bayesian nonlinear approaches to the calibration problem. Racine-Poon (1988) used a Bayesian approach to a nonlinear calibration problem arising from agrochemical soil bioassays (Osborne 1991). Racine-Poon (1988) show that the posterior distribution of an unknown concentration $\eta$ can be calculated by several methods: maximum likelihood; a numerical integration method based on the Gauss quadrature approach of Naylor and Smith (1982); or an approximation based on the Laplace method for integrals (Tierney and Kadane 1986). A noninformative reference prior (Bernardo 1979) approach for the polynomial calibration model is presented by Eno and Ye (2000). Through a second-degree bioassay example presented by Aitchison and Dunsmore (1975), Eno and Ye (2000) derive a reference prior and make posterior inferences about the calibration distribution. In cases when it is not feasible to transform the data to create a straight line, Eno and Ye (2000) show that the inclusion of a quadratic term appropriately adds flexibility to the model. A Bayesian random effects model is proposed by Fong et al. (2012) for the nonlinear calibration problem. Fong et al. (2012) proposed a calibration method that is robust to dependent outliers. They demonstrated the proposed method on data from the HIV Vaccine Trials Network Laboratory and used a normal-mixture model with dependent error terms to model the experimental noise.\\
\indent Graphite furnace atomic absorption spectroscopy (GFAAS)  is an analytical technique for determining trace metal concentrations in different samples. In GFAAS, calibration curves tend to be nonlinear and polynomial regression is used to evaluate unknown sample concentrations. Given a known sample of concentrations ($x_{i}$) and corresponding absorbance values ($y_{i}$) the following relationship can be assumed:
\begin{equation}\label{eq:quad_eq1}
	y_{i} = \beta_{0} + \beta_{1}x_{i}+\beta_{2}x_{i}^{2} + \epsilon_{i}, \hspace{0.9cm}  i = 1, 2, \dots, m,
\end{equation}
where $\beta_{0}$, $\beta_{1}$, $\beta_{2}$ are the unknown parameters and $\epsilon_{i}$ is an independent zero-mean error term with variance $\sigma^2$. Also, an observed absorbance $y_{0}$ corresponding to an unknown concentration $\xi$ can be modeled as
\begin{equation}\label{eq:quad_eq2}
	y_{0} = \beta_{0} + \beta_{1}\xi+\beta_{2}\xi^{2} + \epsilon, 
\end{equation}
where the error term again is assumed to be independent with mean of zero and variance $\sigma^2$. The estimate of the unknown value $\xi$ is calculated as follow:
\begin{equation}\label{eq:quad_model}
\xi^{*} = \frac{- \hat{\beta}_{1} + \sqrt{\hat{\beta}^{2}_{1}-4\hat{\beta}_{2}\left(\hat{\beta}_{0} - y_{0}\right)}}{2\hat{\beta}_{2}},
\end{equation}
where $\hat{\beta}_{0}$, $\hat{\beta}_{1}$ and $\hat{\beta}_{2}$ are the least squares estimates of $\beta_{0}$, $\beta_{1}$ and $\beta_{2}$. The  estimate of $\xi$ is denoted as $\xi^{*}$ and $y_{0}$ is the observed value associated with the unknown concentration $\xi$. Using Equations (\ref{eq:quad_eq1}) - (\ref{eq:quad_model}) Lundberg and de Mar\'{e} (1980) propose a simple interval estimation approach to the spectroscopy calibration problem when there is small measurement error.\\
\indent For sufficiently small values of the measurement variance $\sigma^{2}$, Lundberg and de Mar\'{e} (1980) state that there is a unique and consistent estimate $\xi^{*}$ of $\xi$ and an asymptotic confidence interval of $\xi$ is obtained as 
\begin{equation*}
\xi^{*} \pm t_{\alpha/2,(m+n)} \dot d(\xi^{*})
\end{equation*}
where $d(\xi^{*})$ is chosen as 
\begin{equation*}
d(\xi^{*}) = \frac{s\left[\frac{1}{n} + {\boldsymbol \Xi}^{\prime}({\bf X}^{\prime}{\bf X})^{-1}{\boldsymbol \Xi}\right]^{1/2}}{\left|\sum_{k=1}^{2}k\hat{\beta}_{k}\xi^{*k}\right|},
\end{equation*}
where $s^{2}$ is the residual variance, {\bf X} is the design matrix and 
\[ {\boldsymbol \Xi} = \left[ \begin{array}{c}
1          \\
\xi^{*} \\
\xi^{*2} \end{array} \right].\] 
\indent Fran\c{c}ois et al. (2004)  examined optimal designs for linear and nonlinear calibration models. They show that for the quadratic model given by Equation (\ref{eq:quad_eq1}), the design points should be to the left part of the calibration domain where the calibration curve slope is smaller. Fran\c{c}ois et al. (2004) state that in this area of the domain, the calibration prediction variance is higher and the design points aim then at decreasing the lack of predictive ability of the model in that area. Conversely, the delta method is used to derive an asymptotic confidence interval for $\xi$ by Fran\c{c}ois et al. (2004) as well as Kirkup and Mulholland (2004). The variance in $\xi^{*}$, written as $\sigma^{2}_{\xi^{*}}$, is given by
\begin{equation}
\sigma^{2}_{\xi^{*}} = \left(\frac{\partial \xi^{*}}{\partial y_{0}}\sigma_{y_{0}}\right)^{2} + {\bf d}_{\xi^{*}}^{\prime}{\bf V}{\bf d}_{\xi^{*}},
\end{equation}\label{eq:delta1}
where {\bf V} is the variance-covariance matrix of ${\boldsymbol \beta}$ and
\[ {\bf d}_{\xi^{*}} = \left[ \begin{array}{c}
\frac{\partial \xi^{*}}{\partial \beta_{0}} \\
\frac{\partial \xi^{*}}{\partial \beta_{1}} \\
\frac{\partial \xi^{*}}{\partial \beta_{2}} \end{array} \right],\] 
thus deriving an asympototic confidence interval for $\xi$
\begin{equation}
\xi^{*} \pm z_{1-\alpha/2} \sqrt{\sigma^{2}_{\xi^{*}}}.
\end{equation}\label{eq:delta2}
For an overview of linear and curvilinear calibration methods that commonly use the quadratic calibration model see Merkle (1983); Kirkup and Mulholland (2004); Lavagnini and Magno (2006); Lim and Yun (2010).\\
\indent Weinreb et al. (1990) address nonlinearity in calibration of the advanced very high resolution radiometer (AVHRR). It is assumed that AVHRR can be calibrated by only two points; an internal calibration target (ICT) and space, but Weinreb et al. (1990) state that by not accounting for nonlinearity, errors as large as $2^{\circ}C$ in inferred scene temperatures. The most direct way to handle the nonlinearity would be to use a quadratic calibration equation (Weinreb et al. 1990).\\
\indent Calibrations are never concluded once and for all. Instrument characteristics are altered by time and use, especially in radiometry, and calibration must be viewed as an iterative process as long as the instrument is in use (Cervenka and Massa 1994). Our study is motivated by extending the dynamic linear calibration model of Rivers and Boone (2014) to incorporate a quadratic term in the presence of nonlinearity. In Section 2, we introduce a Bayesian dynamic nonlinear calibration model akin to that of Lundberg and de Mar\'{e} (1980); Weinreb et al. (1990); Eno (1999); Eno and Ye (2000); Fran\c{c}ois et al. (2004); Kirkup and Mulholland (2004); Hibbert (2006); Lavagnini and Magno (2006); Lim and Yun (2010). In Section 3, we demonstrate through a simulation study how the dynamic nonlinear calibration model performs alongside the static estimator given by Equation (\ref{eq:quad_model}) under various noise conditions. In Section 4 the proposed method is applied to a spectroscopy example and a microwave radiometry example. In the first example, the method is used to determine trace amounts of cadmium (Cd) in water samples and for the second example, it is used to estimate a reference temperature given an observed voltage output measure. In Section 5 we conclude with future work and other considerations.

\section{Dynamic Nonlinear Calibration Model}

When collecting laboratory or field measurements for the purpose of calibration, scientist and engineers face a problem when the subsequent stability of the instrument change in relation to time, temperature, pressure, or some other external factors. These changes may cause the instrument readings to drift since an initial calibration thus making it necessary to recalibrate the instrument (Ziemer and Strauss, 1978). We address this problem by developing a dynamic calibration approach that detects changes in the calibration constants in the presense of nonlinearity.\\
\indent Let \{$({\bf X}, {\bf Y}_{t})|t=1, 2, \dots,T$\} be the reference measurement and responses in the calibration experiment at time $t$ and suppose the relationship can be described by
\begin{equation}\label{eq:obs_eq}
	{\bf Y}_{t} = {\bf X}{\boldsymbol \beta_{t}} + {\boldsymbol \epsilon}_{t}, \hspace{0.9cm}  t = 1, 2, \dots, T,
\end{equation}
where ${\bf Y}_{t}$ is a seres of $r-$dimensional vector of responses, ${\bf X}$ is the fixed ($r \times d$) reference design matrix, 
\[ \left[ \begin{array}{ccc}
1         & X_{1}  & X^{2}_{1} \\
1         & X_{2}  & X^{2}_{2} \\
\vdots & \vdots & \vdots\\
1         & X_{r}   & X^{2}_{r} \end{array} \right],\] 
${\boldsymbol \beta}_{t}$,  (i.e. $\beta_{0t}$, $\beta_{1t}$, and $\beta_{2t}$), is a series of $d-$dimensional vectors of unknown dynamic regression parameters, and ${\boldsymbol \epsilon}_{t}$ is   
a $r-$dimensional vector of independently normally distrubuted error terms with mean ${\bf 0}$ and variance-covariance matrix ${\bf E} = \sigma^{2}_{E}{\bf I}$. Equation (\ref{eq:obs_eq}) is known as the observation equation.\\
\indent The evolving relationship between {\bf X} and ${\bf Y}_{t}$ is expressed by the dynamic parameter vector ${\boldsymbol \beta}_{t}$. The evolution in time of the regression parameters is modelled as
\begin{equation}\label{eq:sys_eq}
{\boldsymbol \beta_{t}} = {\boldsymbol \beta_{t-1}} + {\boldsymbol \omega_{t}},  \hspace{0.9cm}  t = 1, 2, \dots, T,
\end{equation}
where ${\boldsymbol \omega_{t}}$ is a $d-$dimensional vector of independently normally distrubuted error terms with mean ${\bf 0}$ and variance-covariance matrix ${\bf W} = \sigma^{2}_{W}{\boldsymbol \Omega}$, and ${\boldsymbol \Omega}=\left[{\bf X}^{\prime}{\bf X}\right]^{-1}$. Equation (\ref{eq:sys_eq}) is known as the system equation.\\
\indent The observation vector ${\bf Y}_{t}$ and the dynamic regression parameter vector ${\boldsymbol \beta_{t}}$ are both random variables, thus the expected values  $\hat{\bf Y}_{t}$ and $\hat{\boldsymbol \beta_{t}}$ are the means of their respective distributions and must be estimated sequentially. The one-step forecast for ${\bf Y}_{t}$ and posterior distributions ${\boldsymbol \beta_{t}}$ for each time $t$ are as follows in Algorithm~\ref{algo:DLM} (Dynamic Linear Regression Models algorithm). See West et al. (1985); West and Harrison (1997) for a more detailed discussion of Dynamic Linear Regression Models (DLRMs).\\
\begin{algorithm}[!bt]
Initialize $t=0$\\
\{Initial information $({\boldsymbol \beta_{0}} | D_{0}) \sim  N_{d}[{\bf m_{0}, C_{0}}]$\}\\
\KwIn{${\bf m_{0}}$, ${\bf C_{0}}$, {\bf E}, {\bf W}}
{\bf loop}\\
\quad$t = t + 1$\\
\quad\{Compute prior at t:  $({\boldsymbol \beta_{t}} | D_{t-1}) \sim  N_{d}[{\bf a_{t}, R_{t}}]$\}\\
\quad\quad ${\bf a}_{t} = {\bf m}_{t-1}$\\
\quad\quad ${\bf R}_{t} = {\bf C}_{t-1} + {\bf W}$\\
\quad\KwIn{{\bf X}}
\quad\{Compute forecast at t:  $({\bf Y_{t}} | D_{t-1}) \sim  N_{r}[{\bf f_{t}, Q_{t}}]$\}\\
\quad\quad ${\bf f}_{t} = {\bf Xa}_{t}$\\
\quad\quad ${\bf Q}_{t} = {\bf X}{\bf R}_{t}{\bf X}^{\prime} + {\bf E}$\\
\quad\KwIn{${\bf Y}_{t}$}
\quad\{Compute forecast error ${\bf e}_{t}$\}\\
\quad\quad ${\bf e}_{t} = {\bf Y}_{t} - {\bf f}_{t}$\\
\quad\{Compute adaptive gain matrix ${\bf A}_{t}$\}\\
\quad\quad ${\bf A}_{t} = {\bf Q}^{-1}_{t}{\bf X}{\bf R}_{t}$\\
\quad\{Compute posterior at t:  $({\boldsymbol \beta_{t}} | D_{t}) \sim  N_{d}[{\bf m}_{t},{\bf C}_{t}]$\}\\
\quad\quad ${\bf m}_{t} = {\bf a}_{t} + {\bf A}_{t}{\bf e}_{t}$\\
\quad\quad ${\bf C}_{t} = {\bf R}_{t} - {\bf A}^{\prime}_{t}{\bf Q}_{t}{\bf A}_{t}$\\
{\bf end loop}
\caption{Updating ({\it DLRM}) Dynamic Linear Regression Model }
\label{algo:DLM}
\end{algorithm}
\indent Furthermore, let $\{y_{0t}|t=1,2,\dots,T\}$ be the observation from the second stage of the calibration experiment corresponding to an unknown reference of interest $x_{0t}$, and 
\begin{equation}\label{eq:calib_eq}
	y_{0t} = \beta_{0t} + \beta_{1t}x_{0t}+\beta_{2t}x_{0t}^{2} + \epsilon_{0t}, \hspace{0.9cm}  t = 1, 2, \dots, T,
\end{equation}
where $\beta_{0t}$, $\beta_{1t}$, and $\beta_{2t}$ are the time dependent regression coefficients and $\epsilon_{0t}$ are assumed to be independently normally distributed with mean 0 and variance $\sigma^{2}_{E}$. Since the quadratic model in Equations (\ref{eq:obs_eq}) and (\ref{eq:sys_eq}) are not monotic on $\mathbb{R}$, the domain will be restricted to where it is strictly increasing.\\
\indent We assume that the first stage of calibration experiment is independent of the second stage, therefore $x_{0t}$ is independent of (${\boldsymbol \beta}_{t}, {\boldsymbol \Gamma}$) where the joint prior distribution is 
\begin{equation}
\pi(x_{0t}, {\boldsymbol \beta}_{t}, {\boldsymbol \Gamma}) = \pi(x_{0t}) \pi({\boldsymbol \beta}_{t}, {\boldsymbol \Gamma})
\end{equation}
and ${\boldsymbol \Gamma}^{\prime} = \left[\sigma^{2}_{E}~~\sigma^{2}_{W}\right]$. The posterior of $(x_{0t}, {\boldsymbol \beta}_{t}, {\boldsymbol \Gamma})$ is then given by 
\begin{eqnarray}\label{eq:calib_post}
\pi(x_{0t}, {\boldsymbol \beta}_{t}, {\boldsymbol \Gamma} | y_{0t}, {\bf Y}_{t}) &\propto & f(y_{0t}, {\bf Y}_{t} | x_{0t}, {\boldsymbol \beta}_{t}, {\boldsymbol \Gamma}) \pi(x_{0t}, {\boldsymbol \beta}_{t}, {\boldsymbol \Gamma})\nonumber\\
                                                                                                    &\propto & f(y_{0t} | x_{0t}, {\boldsymbol \beta}_{t}, {\boldsymbol \Gamma})\pi(x_{0t}) f({\bf Y}_{t} | {\boldsymbol \beta}_{t}, {\boldsymbol \Gamma})\pi({\boldsymbol \beta}_{t}, {\boldsymbol \Gamma})\nonumber\\
                                                                                                    &\propto & f(y_{0t} | x_{0t}, {\boldsymbol \beta}_{t}, {\boldsymbol \Gamma}) \pi({\boldsymbol \beta}_{t}, {\boldsymbol \Gamma} | {\bf Y}_{t})\pi(x_{0t}),
\end{eqnarray} 
where ${\bf Y}_{t}$ and $y_{0t}$ are  respectively the observations from the first and second stages of calibration. Our knowledge about the evolving relationship established in the calibration experiment at each time point is given by the posterior density $\pi({\boldsymbol \beta}_{t}, {\boldsymbol \Gamma} | {\bf Y}_{t})$ which is the middle term in Equation (\ref{eq:calib_post}) and found by Algorithm \ref{algo:DLM} for given values of ${\boldsymbol \Gamma}$. Following Algorithm \ref{algo:DLM} and using multivariate normal thoery (West and Harrison 1997) we have 
\begin{equation*}
\pi({\boldsymbol \beta}_{t} | {\bf Y}_{t}, {\boldsymbol \Gamma}) \sim N_{d}({\bf m}_{t}, {\bf C}_{t}),
\end{equation*}
where ${\bf m}_{t}$ is the posterior mean and ${\bf C}_{t}$ is the variance-covariance matrix of ${\boldsymbol \beta}_{t}$ at time $t$.\\
\begin{algorithm}[!bth]
\begin{enumerate}
	\item Draw {\bf sample} candidates $(\boldsymbol \Theta^{(1)}), \dots, (\boldsymbol \Theta^{(m)})$ {\it i.i.d.} from $g({\boldsymbol  \Theta})$\\
	\item Calculate the standardized {\bf importance} weights, $w({\boldsymbol \Theta}^{(i)}) = \frac{f({\boldsymbol \Theta}^{(i)})/g({\boldsymbol \Theta}^{(i)})}{\sum f({\boldsymbol \Theta}^{(i)})/g({\boldsymbol \Theta}^{(i)})} \mbox{ for } i = 1,\dots,m$\\
	\item {\bf Resample} ${\boldsymbol \Gamma}^{(1)}, \dots, {\boldsymbol \Gamma}^{(n)} \mbox{ from } {\boldsymbol  \Theta}^{(1)}, \dots, {\boldsymbol  \Theta}^{(m)}$ with replacement of probability $w({\boldsymbol  \Theta}^{(1)}), \dots, w({\boldsymbol  \Theta}^{(m)})$ respectively.
\end{enumerate}
\caption{{\sc ({\it SIR})} Sampling Importance Resampling}\label{algo:SIR}
\end{algorithm}
\setlength{\textfloatsep}{5pt}
 \indent The first term in Equation (\ref{eq:calib_post}), $f(y_{0t} | x_{0t}, {\boldsymbol \beta}_{t}, {\boldsymbol \Gamma})$, is the likelihood function for the second stage of the calibration experiment which provides information from the data and $\pi(x_{0t})$ is the prior density for the unknown calibration reference $x_{0t}$. We wish to obtain the conditional posterior density $\pi(x_{0t}| y_{0t}, {\bf Y}_{t})$ at each time $t$. In order to achieve this we will have to integrate over $({\boldsymbol \beta}, {\boldsymbol \Gamma})$.\\
\indent We reduce the parameter space by centering and scaling the data such that 
\begin{equation*}
\sum^{r}_{i=1} x_{i} =0 \mbox{ and } \frac{1}{n}\sum^{r}_{i=1} x^{2}_{i} =1
\end{equation*}
and the intercept term $\beta_{0t}$ is eliminated by moving the origin of the calibration (Hibbert 2006) so the model in Equation (\ref{eq:calib_eq}) is written as
\begin{equation}\label{eq:red_calib_eq}
y_{0t} - \bar{y}_{t} = \beta_{1t}\left(x_{0t}-\bar{x}\right)-\beta_{2t}\left(x_{0t}^{2}-\overline{x^{2}}\right) + \epsilon_{0t}, \hspace{0.9cm}  t = 1, 2, \dots, T,
\end{equation}
where $\bar{x}=\frac{1}{r}\sum_{i=1}^{r}x_{i}$ and $\overline{x^{2}} = \frac{1}{r}\sum_{i=1}^{r}x^{2}_{i}$. In the second stage of the calibration experiment, an observation $y_{0t}$ allows calculation of an unknown reference at time $t$ by:
\begin{equation}\label{eq:dlm_calib2}
\hat{x}_{0t} = \frac{- \hat{\beta}_{1t} \pm \sqrt{\hat{\beta}^{2}_{1t}-4\hat{\beta}_{2t}\left(\bar{y}_{t} - y_{0t} - \hat{\beta}_{1t}\bar{x} - \hat{\beta}_{2t}\overline{x^{2}}\right)}}{2\hat{\beta}_{2t}}.
\end{equation}
The quadratic model in Equation (\ref{eq:dlm_calib2}) has two possible roots on $\mathbb{R}$ but the solution of interest depends on the sign of $\hat{\beta}_{1t}$. The solution is the increasing part of Equation (\ref{eq:dlm_calib2}) when $\hat{\beta}_{1t} < 0$ and the decreasing part of Equation (\ref{eq:dlm_calib2}) is the solution when $\hat{\beta}_{1t} > 0$.\\
\indent Given the reduced parameter space and Equation (\ref{eq:red_calib_eq}), the likelihood function at time $t$ is expressed as
\begin{equation*}
f(y_{0t} | x_{0t}, {\boldsymbol \beta}_{t}, {\boldsymbol \Gamma}) \propto \mbox{exp}\Bigg\{-\frac{1}{2}\left[\sigma^{-2}_{Y_t}(y_{0t} - \bar{y}_{t} - \beta_{1t}(x_{0t}-\bar{x})-\beta_{2t}(x_{0t}^{2}-\overline{x^{2}}))^{2} \right] \Bigg\},
\end{equation*}
the prior density at time $t$ is
\begin{equation*}
\pi(x_{0t}) \propto \mbox{exp}\Bigg\{-\frac{x^{2}_{0t}}{2} \Bigg\},
\end{equation*}
and integrating Equation (\ref{eq:calib_post}) with respect to ${\boldsymbol \beta}_{t}$ for a given ${\boldsymbol \Gamma}$, produces
\begin{eqnarray*}
\pi(x_{0t}| y_{0t}, {\bf Y}_{t}, {\boldsymbol \Gamma}) & \propto & f(y_{0t}|x_{0t},{\boldsymbol \Gamma}, {\bf Y}_{t})\pi(x_{0t})\nonumber\\ 
								     & \propto &  \mbox{exp}\Bigg\{-\frac{1}{2}\left[\sigma^{-2}_{Y_t}(y_{0t} - \bar{y}_{t} - \beta_{1t}(x_{0t}-\bar{x})-\beta_{2t}(x_{0t}^{2}-\overline{x^{2}}))^{2} \right] -\frac{1}{2}x^{2}_{0t}  \Bigg\}\nonumber\\
                                                                                          & \propto &  \mbox{exp}\Bigg\{-\frac{1}{2}\left[\sigma^{-2}_{Y_t}(\hat{x}_{0t} - x_{0t})^{2} \right] -\frac{1}{2}x^{2}_{0t} \Bigg\}.
\end{eqnarray*}
By completing the square, the posterior density for the unknown reference measurement at time time $t$ is
\begin{equation}\label{eq:post_draw}
\pi(x_{0t}| y_{0t}, {\bf Y}_{t}, {\boldsymbol \Gamma}) \sim N(\mu_{x_{0t}}, \sigma^{2}_{x_{0t}}),
\end{equation}
with
\begin{eqnarray*}
\mu_{x_{0t}} &=& \frac{x_{0t}}{1+\sigma^{2}_{Y_t}},\\
\sigma^{2}_{x_{0t}} &=& \frac{1}{1+\sigma^{2}_{Y_t}},
\end{eqnarray*}
and 
\begin{equation*}
\sigma^{2}_{Y_t} = \mbox{tr}({\bf Q}_{t})
\end{equation*}
where tr(~.~) denotes trace of the one-step forecast variance-covariance matrix in Algorithm \ref{algo:DLM}.\\
\indent To obtain posterior samples of Equation (\ref{eq:post_draw}) for a given ${\boldsymbol \Gamma}$ we use the Sampling Importance Resampling (SIR) algorithm (Rubin 1987; Smith and Gelfand 1992; Givens and Hoeting 2005; Albert 2007) to draw random samples from the target distribution $\pi(x_{0t}| y_{0t}, {\bf Y}_{t}, {\boldsymbol \Gamma})$ by using a candidate distribution $g({\boldsymbol  \Theta})$ for ${\boldsymbol \Gamma}$, thus deriving the posterior densities of interest. Algorithm~\ref{algo:SIR} is the Sampling Importance Resampling algorithm.\\ 
\indent For most applications, it is believed {\it a priori} that the observational variance $\sigma^{2}_{E}$ is greater in magnitude than the system variance $\sigma^{2}_{W}$, such that
\begin{equation*}
\sigma^{2}_{W} < \sigma^{2}_{E}.
\end{equation*}
To enforce this belief about the variance relationship we utilize the following prior distributions:
\begin{eqnarray}
\sigma^{2}_{E} &\sim& Uniform(0,\alpha_{E}) \label{eq:prior1}\\
\sigma^{2}_{W}|\sigma^{2}_{E} &\sim& Uniform(0, \sigma^{2}_{E}), \label{eq:prior2}
\end{eqnarray}
where prior distributions (\ref{eq:prior1}) and (\ref{eq:prior2}) ensure the system variance to be less than the observation variance. Since these are proper prior distributions the resulting posterior distribution will also be proper. We combine prior distributions (\ref{eq:prior1}), (\ref{eq:prior2}), Algorithm \ref{algo:DLM} (DLM), Equation (\ref{eq:post_draw}), and Algorithm \ref{algo:SIR} (SIR) together and propose the Dynamic Calibration Method in Algorithm \ref{algo:DynCal}. 
\vfill
\scalebox{0.90}{
\begin{algorithm}[H] 
	\begin{enumerate}
	\item Draw $M$ {\it i.i.d.} {\bf sample} candidates for $(\sigma^{2}_{E}, \sigma^{2}_{W})$ from $g({\boldsymbol \Theta})$. 
	\item Calibration data are fit using the DLRM framework for each of the $M$ proposal samples $(\sigma^{2(m)}_{E}, \sigma^{2(m)}_{W})$, with the prior moments for $({\boldsymbol \beta_{0}} | D_{0})$ as {$\bf m_{0} = 1_{d}$} and {$\bf C_{0} = 100I_{(d \times d)}$}, where ${\bf 1_{d}}$ is a $d-$dimensional vector of ones:	
		\begin{enumerate}[a.]
			\item Data are scaled and centered such that $\sum^{r}_{i=1} x_{i} =0$,  $\frac{1}{n}\sum^{r}_{i=1} x^{2}_{i} =1$ and $\beta_{0t} = 0$;
			\item Estimate ${\boldsymbol \beta}^{(m)}_{t}|\sigma^{2(m)}_{E}, \sigma^{2(m)}_{W}$ for the $m^{th}$ proposal sample ${\boldsymbol \Theta}^{(m)}$ for all time $t$;
			\item Draw a sample from $\pi(x_{0t}| y_{0t}, {\bf Y}_{t}, {\boldsymbol \Gamma})$ given the $m^{th}$ proposal sample ${\boldsymbol \Theta}^{(m)}$ at time $t$; 
			\item Calculate log-likelihoods, $log[f({\boldsymbol \Theta}^{(m)})]$, for each $(\sigma^{2(m)}_{E_t}, \sigma^{2(m)}_{W_t})$ pair.
		\end{enumerate}
		\item Sampling Importance Resampling (SIR) is used to simulate  samples of $x_{0t}| {\boldsymbol \beta}_{t},\sigma^{2}_{E}, \sigma^{2}_{W}$ by accepting a subset of $N$ from the proposal density to be distributed according to the posterior density $\pi({\boldsymbol \Gamma}|{\bf Y}_{t})$ with candidate density $g({\boldsymbol \Theta})$.
		\begin{enumerate}[a.]
			\item Calculate the standardized {\bf importance} weights, $w({{\boldsymbol \Theta}^{(1)}}), \dots, w({{\boldsymbol \Theta}^{(M)}})$ , where $w({{\boldsymbol \Theta^{(m)}}}) = log[f({\boldsymbol \Theta}^{(m)})] - log[g({\boldsymbol \Theta}^{(m)})]$ for the $m^{th}$ proposal sample;
			\item  {\bf Resample} $N$ calibrated time series from the $M$ proposal values with replacement given probabilities $p({{\boldsymbol \Theta}^{(m)}})$ where 
		\begin{equation*}
			p({{\boldsymbol \Theta}^{(m)}}) = \frac{e^{w({{\boldsymbol \Theta^{(m)}}})}}{\sum_{j=1}^{M}e^{w({{\boldsymbol \Theta^{(j)}}})}}.
		\end{equation*}
		\end{enumerate}
	\item Rescale calibrated time series to original scale and record summary statistics (i.e. medians and credible sets) across each time $t$ .
\end{enumerate}
\caption{Dynamic~Calibration\label{algo:DynCal}}
\end{algorithm}
}

\section{Simulation Study}\label{sec:Sim_Study}

\begin{figure}[ht]
\centering
{\includegraphics[width=5in,height=2.25in]{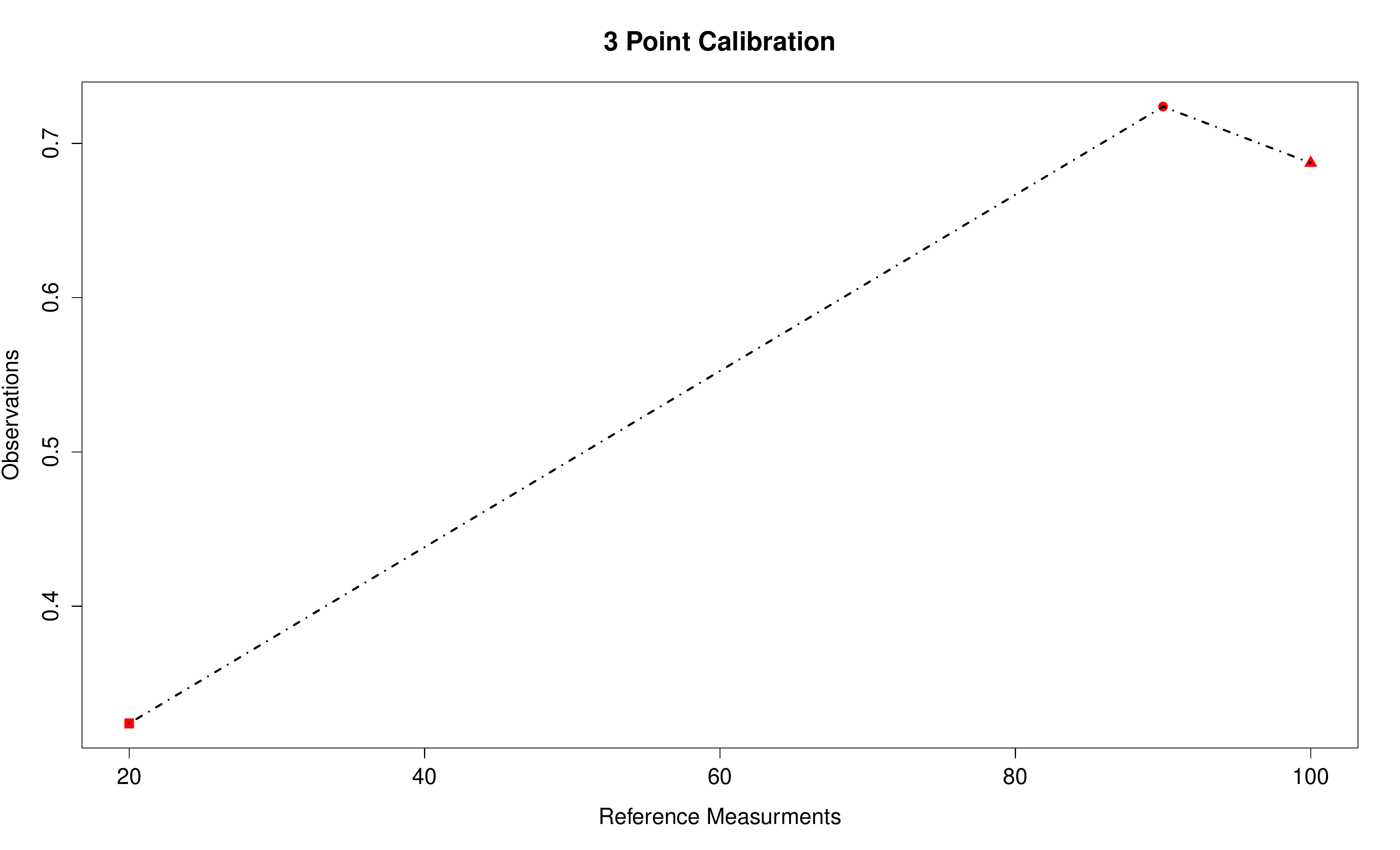}
	\subcaption{3-Point Calibration}} 
~\quad
{\includegraphics[width=5in,height=2.25in]{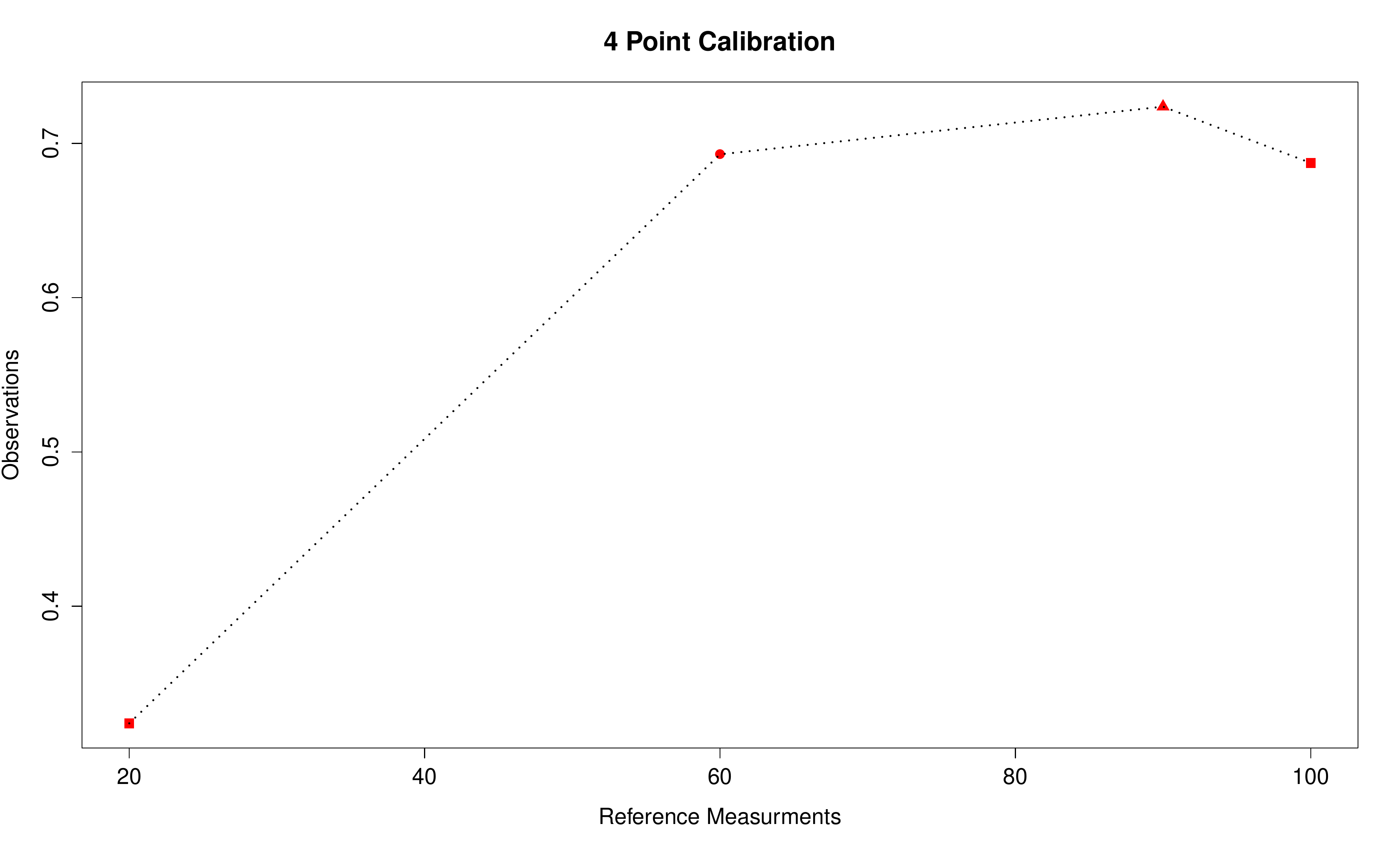}
	\subcaption{4-Point Calibration}}
~
{\includegraphics[width=5in,height=2.25in]{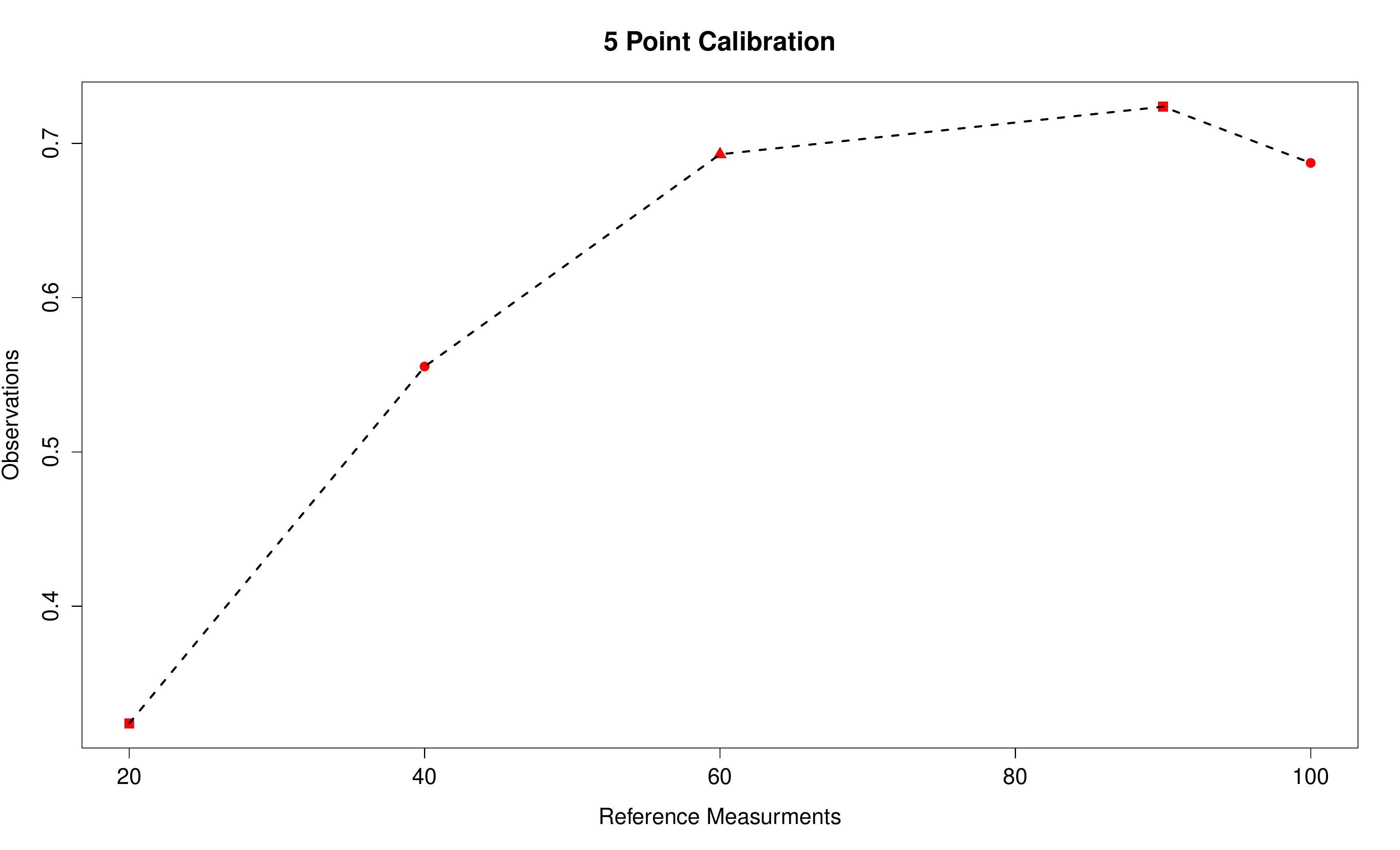}
	\subcaption{5-Point Calibration}}
     \caption{Three reference design schemes: (a) 3-point calibration; (b) 4-point calibration; (c) 5-point calibration}\label{fig:cali_scheme}
\end{figure}

In this section, we conducted several simulation experiments to illustrate the adaptability  of the proposed method. In the simulation, we consider nine cases based on variance combinations for the observation and system variances, $(\sigma^{2}_{E}, \sigma^{2}_{W})$, for the Dynamic Linear Models.
The true values for $\sigma^{2}_{E}$ and $\sigma^{2}_{W}$ used in the simulation study are $(0.00001, 0.0001, 0.001)$ and $(0.00005, 0.0001, 0.001)$, respectively. For each variance pair $(\sigma^{2}_{E}, \sigma^{2}_{W})$, the number of simulated realizations is $N=100$. In each realization, the number of simulated time periods is $T=1000$. The posterior densities samples drawn from $\pi(x_{0t}| y_{0t}, {\bf Y}_{t}, {\boldsymbol \Gamma})$ by the Dynamic Calibration Approach (Algorithm \ref{algo:DynCal}) is assessed via the square root of the averaged mean squared errors ($RAMSE$), the average interval width ($AvIW$), and the average coverage probability ($AvCP$):
\begin{equation}
RAMSE = \left[\frac{1}{N}\sum_{j=1}^{N}MSE_{j}\right]^{\frac{1}{2}},\nonumber
\end{equation}
where
\begin{equation}
MSE_{j} = \frac{1}{T}\sum_{t=1}^{T}(\hat{x}_{0t}-x_{0t})^{2};\nonumber
\end{equation}
and
\begin{equation*}
AvIW = \frac{1}{N}\sum_{j=1}^{N}IW_{j}.
\end{equation*}
where
\begin{equation*}
IW_{j} = \frac{1}{T}\sum_{t=1}^{T}(x_{0t}^{U} - x_{0t}^{L}).
\end{equation*}
Note that if  $x_{0t}^{L}$ is the 0.025 posterior quantile for  $x_{0t}$, and  $x_{0t}^{U}$ is the 0.975 posterior quantile for $x_{0t}$,  where $x_{0t}$ is the true value of the calibration target from the second stage of experimentation, then  $(x_{0t}^{L}, x_{0t}^{U})$ is a $95\%$ credible interval.\\
\indent Using the credible interval above we defined the coveraged probability ($CP$) which is calculated as such
\begin{equation}
 CP_{j} = \frac{1}{T}\sum_{t=1}^{T}\psi_{t}\nonumber
\end{equation}
where
\[ \psi_{t} = P[x_{0t}^{L} < x_{0t} < x_{0t}^{U}] = \left\{ \begin{array}{ll}
         0 & \mbox{if $x_{0t} \not\in (x_{0t}^{L}, x_{0t}^{U})$};\\
	 &\\
         1 & \mbox{if $x_{0t} \in (x_{0t}^{L}, x_{0t}^{U})$}.\end{array} \right. \]
The average coverage probability ($AvCP$) is calculated by averaging across the number of replications in the simulation study, where
\begin{equation}
 AvCP = \frac{1}{N}\sum_{j=1}^{N}CP_{j}.\nonumber
\end{equation}
\indent The data are generated from the following model:
\begin{equation}\label{eq:sim_model}
{\bf Y}_{t} = {\bf X}{\boldsymbol \beta}_{t} + {\boldsymbol \epsilon}_{t},
\end{equation}
where ${\bf X}$ is a known fixed model matrix of reference values augmented with a column of 1's. The dynamic Bayesian nonlinear calibration model will be assessed across three different reference measurement schemes. In the first scheme, the reference measurements will be taken at [20,~90,~100]. The references will be placed at [20,~60,~90,~100] for the second scheme and at [20,~40,~60,~90,~100] for the final and third scheme. See Figure \ref{fig:cali_scheme}. \\
 \indent The vector of regression parameters, ${\boldsymbol \beta}_{t}$, are random draws from a multivariate normal distribution with mean vector $[ \beta_{0} ~ \beta_{1}~ \beta_{2}]^{'}$ and variance-covariance matrix, ${\bf W} = \sigma^{2}_{W}\left[{\bf X}^{'} {\bf X}\right]^{-1}$ for $t = 1, \dots, T$, were $\beta_{0} = -0.0007$, $\beta_{1} = 0.01858$, and $\beta_{1} = -0.000117$. For each $t$, the random multivariate error vector is
\begin{equation}
 {\boldsymbol \epsilon}_{t} \sim N_{r}[{\bf 0}, \sigma^{2}_{E}{\bf I}].
\end{equation}
\indent Tables \ref{tab:sim_results3} - \ref{tab:sim_results5} summarizes the results of the dynamic nonlinear calibration estimator under different variance pairs, $(\sigma^{2}_{E}, \sigma^{2}_{W})$, and results for the static model given by Equation (\ref{eq:quad_model}). An asymptotic variance and $95\%$ confidence interval for the static quadratic calibration is calculated from Equations (\ref{eq:delta1}) and (\ref{eq:delta2}).  The dynamic calibration approach is denoted as DC and the static calibration approach is denoted as SC in the tables. From Tables \ref{tab:sim_results3} - \ref{tab:sim_results5}, we make the following observations:
\begin{itemize}
	\item In Table \ref{tab:sim_results3}, the 3 reference case, the static method require that the degrees of freedom $n-3$ be $\geq 1$ in order to estimate the variance $\sigma^{2}$. The dynamic calibration method is not dependent upon the degrees of freedom for estimation of the error variance, thus, credible intervals and interval widths can be computed.
	\item The RAMSE for the dynamic method is $11\%$ to $41\%$ smaller than the the RAMSE for the static method.
	\item The RAMSEs of the dynamic estimator and the static estimator increase as the observation variance $\sigma^{2}_{E}$ increases.
	\item For the 4 and 5 reference models, when $\sigma^{2}_{E}=0.001$ the average coverage propability (ACP) is just barely above $40\%$.
	\item Given $\sigma^{2}_{E}$, the RAMSEs increase slowly as $\sigma^{2}_{W}$ increases.
	\item For the 4 and 5 reference models, the average interval width (AIW) increases for the dynamic method as the observation variance $\sigma^{2}_{E}$ increase. The interval widths for the static calibration method is consistent regardless of the observational or system noise.
\end{itemize}
%
\begin{table}[]
\centering
\captionof{table}[Summary of the $RAMSE$, $AIW$ and $AvCP$ with 3 references for the dynamic nonlinear calibration estimator (DC) and the static quadratic estimator (SC) under $(\sigma^{2}_{E}, \sigma^{2}_{W})$.]{
Summary of the $RAMSE$, $AIW$ and $AvCP$ with 3 references for the dynamic nonlinear calibration estimator (DC) and the static quadratic estimator (SC) under $(\sigma^{2}_{E}, \sigma^{2}_{W})$. 
}
\begin{tabular}{rcccccc}
\toprule
&\multicolumn{6}{c}{{\bf 3 Reference Model}}\\
\midrule
&\multicolumn{6}{c}{$\sigma^{2}_{W}=0.00005$}\\
& \multicolumn{2}{c}{{{\bf RAMSE}}}
& \multicolumn{2}{c}{{{\bf AIW}}}
& \multicolumn{2}{c}{{{\bf ACP}}}\\
\cmidrule(lr{0.125em}){2-3}%
\cmidrule(lr{0.125em}){4-5}%
\cmidrule(lr{0.125em}){6-7}%
$\sigma^{2}_{E}$ & DC & SC & DC & SC & DC & SC \\ 
0.00001 & 0.249 & 0.382 & 1.471 & N/A & 0.992 & N/A \\ 
  0.0001 & 0.643 & 1.099 & 4.642 & N/A & 0.997 & N/A \\ 
    0.001 & 1.988 & 3.391 & 14.592 & N/A & 0.997 & N/A \\
\midrule
&\multicolumn{6}{c}{$\sigma^{2}_{W}=0.0001$}\\
& \multicolumn{2}{c}{{{\bf RAMSE}}}
& \multicolumn{2}{c}{{{\bf AIW}}}
& \multicolumn{2}{c}{{{\bf ACP}}}\\
\cmidrule(lr{0.125em}){2-3}%
\cmidrule(lr{0.125em}){4-5}%
\cmidrule(lr{0.125em}){6-7}%
 $\sigma^{2}_{E}$ & DC & SC & DC & SC & DC & SC \\ 
0.00001 & 0.291 & 0.413 & 1.473 & N/A & 0.984 & N/A \\ 
  0.0001 & 0.661 & 1.109 & 4.645 & N/A & 0.996 & N/A \\ 
    0.001 & 1.999 & 3.399 & 14.592 & N/A & 0.997 & N/A \\ 
\midrule
&\multicolumn{6}{c}{$\sigma^{2}_{W}=0.001$}\\
& \multicolumn{2}{c}{{{\bf RAMSE}}}
& \multicolumn{2}{c}{{{\bf AIW}}}
& \multicolumn{2}{c}{{{\bf ACP}}}\\
\cmidrule(lr{0.125em}){2-3}%
\cmidrule(lr{0.125em}){4-5}%
\cmidrule(lr{0.125em}){6-7}%
$\sigma^{2}_{E}$ & DC & SC & DC & SC & DC & SC \\ 
0.00001 & 0.716 & 0.806 & 1.500 & N/A & 0.688 &  N/A\\ 
  0.0001 & 0.936 & 1.306 & 4.652 & N/A & 0.983 &  N/A\\ 
    0.001 & 2.126 & 3.475 & 14.600 & N/A & 0.996 & N/A \\ 
   \hline
\end{tabular}\label{tab:sim_results3}
\end{table}
%

\clearpage

\begin{table}[]
\centering
\captionof{table}[Summary of the $RAMSE$, $AIW$ and $AvCP$ with 4 references for the dynamic nonlinear calibration estimator (DC) and the static quadratic estimator (SC) under $(\sigma^{2}_{E}, \sigma^{2}_{W})$.]{
Summary of the $RAMSE$, $AIW$ and $AvCP$ with 4 references for the dynamic nonlinear calibration estimator (DC) and the static quadratic estimator (SC) under $(\sigma^{2}_{E}, \sigma^{2}_{W})$.
}
\begin{tabular}{rcccccc}
\toprule
&\multicolumn{6}{c}{{\bf 4 Reference Model}}\\
\midrule
&\multicolumn{6}{c}{$\sigma^{2}_{W}=0.00005$}\\
& \multicolumn{2}{c}{{{\bf RAMSE}}}
& \multicolumn{2}{c}{{{\bf AIW}}}
& \multicolumn{2}{c}{{{\bf ACP}}}\\
\cmidrule(lr{0.125em}){2-3}%
\cmidrule(lr{0.125em}){4-5}%
\cmidrule(lr{0.125em}){6-7}%
$\sigma^{2}_{E}$ & DC & SC & DC & SC & DC & SC \\ 
0.00001 & 0.247 & 0.358 & 1.500 & 3.763 & 0.993 & 0.960 \\ 
  0.0001 & 0.739 & 1.072 & 4.741 & 3.920 & 0.994 & 0.931 \\
  0.001  & 2.332 & 3.376 & 14.875 & 3.920 & 0.995 & 0.451 \\
\midrule
&\multicolumn{6}{c}{$\sigma^{2}_{W}=0.0001$}\\
& \multicolumn{2}{c}{{{\bf RAMSE}}}
& \multicolumn{2}{c}{{{\bf AIW}}}
& \multicolumn{2}{c}{{{\bf ACP}}}\\
\cmidrule(lr{0.125em}){2-3}%
\cmidrule(lr{0.125em}){4-5}%
\cmidrule(lr{0.125em}){6-7}%
$\sigma^{2}_{E}$ & DC & SC & DC & SC & DC & SC \\ 
0.00001 & 0.259 & 0.365 & 1.500 & 3.763 & 0.991 & 0.960 \\ 
0.0001   & 0.747 & 1.081 & 4.742 & 3.920 & 0.994 & 0.930 \\ 
0.001     & 2.336 & 3.379 & 14.875 & 3.920 & 0.994 & 0.451 \\
\midrule
&\multicolumn{6}{c}{$\sigma^{2}_{W}=0.001$}\\
& \multicolumn{2}{c}{{{\bf RAMSE}}}
& \multicolumn{2}{c}{{{\bf AIW}}}
& \multicolumn{2}{c}{{{\bf ACP}}}\\
\cmidrule(lr{0.125em}){2-3}%
\cmidrule(lr{0.125em}){4-5}%
\cmidrule(lr{0.125em}){6-7}%
$\sigma^{2}_{E}$ & DC & SC & DC & SC & DC & SC \\ 
0.00001 & 0.427 & 0.498 & 1.502 & 3.920 & 0.917 & 0.998 \\ 
0.0001   & 0.825 & 1.131 & 4.742 & 3.920 & 0.990 & 0.924 \\ 
0.001     & 2.380 & 3.411 & 14.876 & 3.920 & 0.994 & 0.446 \\
   \hline
\end{tabular}\label{tab:sim_results4}
\end{table}
%
%
%

\clearpage

\begin{table}[]
\centering
\captionof{table}[Summary of the $RAMSE$, $AIW$ and $AvCP$ with 5 references for the dynamic nonlinear calibration estimator (DC) and the static quadratic estimator (SC) under $(\sigma^{2}_{E}, \sigma^{2}_{W})$.]{
Summary of the $RAMSE$, $AIW$ and $AvCP$ with 5 references for the dynamic nonlinear calibration estimator (DC) and the static quadratic estimator (SC) under $(\sigma^{2}_{E}, \sigma^{2}_{W})$.
}
\begin{tabular}{rcccccc}
\toprule
&\multicolumn{6}{c}{{\bf 5 Reference Model}}\\
\midrule
&\multicolumn{6}{c}{$\sigma^{2}_{W}=0.00005$}\\
& \multicolumn{2}{c}{{{\bf RAMSE}}}
& \multicolumn{2}{c}{{{\bf AIW}}}
& \multicolumn{2}{c}{{{\bf ACP}}}\\
\cmidrule(lr{0.125em}){2-3}%
\cmidrule(lr{0.125em}){4-5}%
\cmidrule(lr{0.125em}){6-7}%
$\sigma^{2}_{E}$ & DC & SC & DC & SC & DC & SC \\ 
0.00001 & 0.249 & 0.350 & 1.569 & 3.920 & 0.994 & 1.000 \\ 
  0.0001 & 0.768 & 1.083 & 4.955 & 3.920 & 0.995 & 0.927 \\ 
  0.001   & 2.429 & 3.374 & 15.515 & 3.920 & 0.995 & 0.452 \\
\midrule
&\multicolumn{6}{c}{$\sigma^{2}_{W}=0.0001$}\\
& \multicolumn{2}{c}{{{\bf RAMSE}}}
& \multicolumn{2}{c}{{{\bf AIW}}}
& \multicolumn{2}{c}{{{\bf ACP}}}\\
\cmidrule(lr{0.125em}){2-3}%
\cmidrule(lr{0.125em}){4-5}%
\cmidrule(lr{0.125em}){6-7}%
$\sigma^{2}_{E}$ & DC & SC & DC & SC & DC & SC \\ 
0.00001  & 0.257 & 0.356 & 1.569 & 3.920 & 0.993 & 1.000 \\ 
  0.0001  & 0.770 & 1.082 & 4.955 & 3.920 & 0.995 & 0.929 \\ 
  0.001    & 2.432 & 3.377 & 15.515 & 3.878 & 0.995 & 0.447 \\
\midrule
&\multicolumn{6}{c}{$\sigma^{2}_{W}=0.001$}\\
& \multicolumn{2}{c}{{{\bf RAMSE}}}
& \multicolumn{2}{c}{{{\bf AIW}}}
& \multicolumn{2}{c}{{{\bf ACP}}}\\
\cmidrule(lr{0.125em}){2-3}%
\cmidrule(lr{0.125em}){4-5}%
\cmidrule(lr{0.125em}){6-7}%
$\sigma^{2}_{E}$ & DC & SC & DC & SC & DC & SC \\ 
0.00001  & 0.370 & 0.445 & 1.570 & 3.920 & 0.958 & 0.999 \\ 
 0.0001   & 0.821 & 1.120 & 4.956 & 3.920 & 0.993 & 0.924 \\ 
0.001      & 2.462 & 3.404 & 15.516 & 3.920 & 0.994 & 0.448 \\ 
   \hline
\end{tabular}\label{tab:sim_results5}
\end{table}

\newpage

	\subsection{Simulation with Random Shock}

\begin{table}[]
\centering
\captionof{table}[Summary of the $RAMSE$, $AIW$ and $AvCP$ with 3 references for the dynamic nonlinear calibration estimator (DC) and the static quadratic estimator (SC) under $(\sigma^{2}_{E}, \sigma^{2}_{W})$ with random shocks in system.]{
Summary of the $RAMSE$, $AIW$ and $AvCP$ with 3 references for the dynamic nonlinear calibration estimator (DC) and the static quadratic estimator (SC) under $(\sigma^{2}_{E}, \sigma^{2}_{W})$ with random shocks in system. 
}
\begin{tabular}{rcccccc}
\toprule
&\multicolumn{6}{c}{{\bf 3 Reference Model}}\\
\midrule
&\multicolumn{6}{c}{$\sigma^{2}_{W}=0.00005$}\\
& \multicolumn{2}{c}{{{\bf RAMSE}}}
& \multicolumn{2}{c}{{{\bf AIW}}}
& \multicolumn{2}{c}{{{\bf ACP}}}\\
\cmidrule(lr{0.125em}){2-3}%
\cmidrule(lr{0.125em}){4-5}%
\cmidrule(lr{0.125em}){6-7}%
$\sigma^{2}_{E}$ & DC & SC & DC & SC & DC & SC \\ 
0.00001 & 2.554 & 2.603 & 1.643 & N/A & 0.953 & N/A \\ 
  0.0001 & 2.587 & 2.788 & 4.770 & N/A & 0.958 & N/A \\ 
    0.001 & 3.115 & 4.234 & 14.652 & N/A & 0.966 & N/A \\
\midrule
&\multicolumn{6}{c}{$\sigma^{2}_{W}=0.0001$}\\
& \multicolumn{2}{c}{{{\bf RAMSE}}}
& \multicolumn{2}{c}{{{\bf AIW}}}
& \multicolumn{2}{c}{{{\bf ACP}}}\\
\cmidrule(lr{0.125em}){2-3}%
\cmidrule(lr{0.125em}){4-5}%
\cmidrule(lr{0.125em}){6-7}%
 $\sigma^{2}_{E}$ & DC & SC & DC & SC & DC & SC \\ 
0.00001 & 2.556 & 2.605 & 1.644 & N/A & 0.945 & N/A \\ 
  0.0001 & 2.589 & 2.790 & 4.770 & N/A & 0.958 & N/A \\ 
    0.001 & 3.120 & 4.239 & 14.653 & N/A & 0.966 & N/A \\ 
\midrule
&\multicolumn{6}{c}{$\sigma^{2}_{W}=0.001$}\\
& \multicolumn{2}{c}{{{\bf RAMSE}}}
& \multicolumn{2}{c}{{{\bf AIW}}}
& \multicolumn{2}{c}{{{\bf ACP}}}\\
\cmidrule(lr{0.125em}){2-3}%
\cmidrule(lr{0.125em}){4-5}%
\cmidrule(lr{0.125em}){6-7}%
$\sigma^{2}_{E}$ & DC & SC & DC & SC & DC & SC \\ 
0.00001 & 2.619 & 2.674 & 1.672 & N/A & 0.674 &  N/A\\ 
  0.0001 & 2.654 & 2.855 & 4.778 & N/A & 0.945 &  N/A\\ 
    0.001 & 3.192 & 4.296 & 14.656 & N/A & 0.962 & N/A \\ 
   \hline
\end{tabular}\label{tab:sim_results3a}
\end{table}	
\clearpage

\begin{table}[]
\centering
\captionof{table}[Summary of the $RAMSE$, $AIW$ and $AvCP$ with 4 references for the dynamic nonlinear calibration estimator (DC) and the static quadratic estimator (SC) under $(\sigma^{2}_{E}, \sigma^{2}_{W})$ with random shocks in system.]{
Summary of the $RAMSE$, $AIW$ and $AvCP$ with 4 references for the dynamic nonlinear calibration estimator (DC) and the static quadratic estimator (SC) under $(\sigma^{2}_{E}, \sigma^{2}_{W})$ with random shocks in system.
}
\begin{tabular}{rcccccc}
\toprule
&\multicolumn{6}{c}{{\bf 4 Reference Model}}\\
\midrule
&\multicolumn{6}{c}{$\sigma^{2}_{W}=0.00005$}\\
& \multicolumn{2}{c}{{{\bf RAMSE}}}
& \multicolumn{2}{c}{{{\bf AIW}}}
& \multicolumn{2}{c}{{{\bf ACP}}}\\
\cmidrule(lr{0.125em}){2-3}%
\cmidrule(lr{0.125em}){4-5}%
\cmidrule(lr{0.125em}){6-7}%
$\sigma^{2}_{E}$ & DC & SC & DC & SC & DC & SC \\ 
0.00001 & 2.506 & 2.604 & 1.641   & 3.920 & 0.953 & 0.960 \\ 
  0.0001 & 2.562 & 2.789 & 4.844   & 3.920 & 0.956 & 0.889 \\
  0.001   & 3.301 & 4.231 & 14.921 & 3.920 & 0.967 & 0.429 \\
\midrule
&\multicolumn{6}{c}{$\sigma^{2}_{W}=0.0001$}\\
& \multicolumn{2}{c}{{{\bf RAMSE}}}
& \multicolumn{2}{c}{{{\bf AIW}}}
& \multicolumn{2}{c}{{{\bf ACP}}}\\
\cmidrule(lr{0.125em}){2-3}%
\cmidrule(lr{0.125em}){4-5}%
\cmidrule(lr{0.125em}){6-7}%
$\sigma^{2}_{E}$ & DC & SC & DC & SC & DC & SC \\ 
0.00001 & 2.507 & 2.605 & 1.640 & 3.920 & 0.951 & 0.960 \\ 
0.0001   & 2.563 & 2.790 & 4.844 & 3.920 & 0.956 & 0.889 \\ 
0.001     & 3.303 & 4.234 & 14.921 & 3.920 & 0.967 & 0.429 \\
\midrule
&\multicolumn{6}{c}{$\sigma^{2}_{W}=0.001$}\\
& \multicolumn{2}{c}{{{\bf RAMSE}}}
& \multicolumn{2}{c}{{{\bf AIW}}}
& \multicolumn{2}{c}{{{\bf ACP}}}\\
\cmidrule(lr{0.125em}){2-3}%
\cmidrule(lr{0.125em}){4-5}%
\cmidrule(lr{0.125em}){6-7}%
$\sigma^{2}_{E}$ & DC & SC & DC & SC & DC & SC \\ 
0.00001 & 2.524 & 2.621 & 1.642 & 3.920 & 0.880 & 0.959 \\ 
0.0001   & 2.582 & 2.807 & 4.845 & 3.920 & 0.952 & 0.882 \\ 
0.001     & 3.330 & 4.259 & 14.922 & 3.920 & 0.966 & 0.425 \\
   \hline
\end{tabular}\label{tab:sim_results4a}
\end{table}
\clearpage
\begin{table}[]
\centering
\captionof{table}[Summary of the $RAMSE$, $AIW$ and $AvCP$ with 5 references for the dynamic nonlinear calibration estimator (DC) and the static quadratic estimator (SC) under $(\sigma^{2}_{E}, \sigma^{2}_{W})$ with random shocks in system.]{
Summary of the $RAMSE$, $AIW$ and $AvCP$ with 5 references for the dynamic nonlinear calibration estimator (DC) and the static quadratic estimator (SC) under $(\sigma^{2}_{E}, \sigma^{2}_{W})$ with random shocks in system.
}
\begin{tabular}{rcccccc}
\toprule
&\multicolumn{6}{c}{{\bf 5 Reference Model}}\\
\midrule
&\multicolumn{6}{c}{$\sigma^{2}_{W}=0.00005$}\\
& \multicolumn{2}{c}{{{\bf RAMSE}}}
& \multicolumn{2}{c}{{{\bf AIW}}}
& \multicolumn{2}{c}{{{\bf ACP}}}\\
\cmidrule(lr{0.125em}){2-3}%
\cmidrule(lr{0.125em}){4-5}%
\cmidrule(lr{0.125em}){6-7}%
$\sigma^{2}_{E}$ & DC & SC & DC & SC & DC & SC \\ 
0.00001 & 0.249 & 0.350 & 1.569 & 3.920 & 0.994 & 1.000 \\ 
  0.0001 & 0.768 & 1.083 & 4.955 & 3.920 & 0.995 & 0.927 \\ 
  0.001   & 2.429 & 3.374 & 15.515 & 3.920 & 0.995 & 0.452 \\
\midrule
&\multicolumn{6}{c}{$\sigma^{2}_{W}=0.0001$}\\
& \multicolumn{2}{c}{{{\bf RAMSE}}}
& \multicolumn{2}{c}{{{\bf AIW}}}
& \multicolumn{2}{c}{{{\bf ACP}}}\\
\cmidrule(lr{0.125em}){2-3}%
\cmidrule(lr{0.125em}){4-5}%
\cmidrule(lr{0.125em}){6-7}%
$\sigma^{2}_{E}$ & DC & SC & DC & SC & DC & SC \\ 
0.00001  & 0.257 & 0.356 & 1.569 & 3.920 & 0.993 & 1.000 \\ 
  0.0001  & 0.770 & 1.082 & 4.955 & 3.920 & 0.995 & 0.929 \\ 
  0.001    & 2.432 & 3.377 & 15.515 & 3.878 & 0.995 & 0.447 \\
\midrule
&\multicolumn{6}{c}{$\sigma^{2}_{W}=0.001$}\\
& \multicolumn{2}{c}{{{\bf RAMSE}}}
& \multicolumn{2}{c}{{{\bf AIW}}}
& \multicolumn{2}{c}{{{\bf ACP}}}\\
\cmidrule(lr{0.125em}){2-3}%
\cmidrule(lr{0.125em}){4-5}%
\cmidrule(lr{0.125em}){6-7}%
$\sigma^{2}_{E}$ & DC & SC & DC & SC & DC & SC \\ 
0.00001  & 0.370 & 0.445 & 1.570 & 3.920 & 0.958 & 0.999 \\ 
 0.0001   & 0.821 & 1.120 & 4.956 & 3.920 & 0.993 & 0.924 \\ 
0.001      & 2.462 & 3.404 & 15.516 & 3.920 & 0.994 & 0.448 \\ 
   \hline
\end{tabular}\label{tab:sim_results5a}
\end{table}

%

	\section{Applications}\label{sec:application}
	\subsection*{Example 1}

As a demonstration of the dynamic nonlinear calibration method, we extend the example of nonlinear calibration presented by Lundberg and de Mar\'{e} (1980).
\begin{figure}[htbp]
\begin{center}
\includegraphics[width = 14cm, height = 8cm]{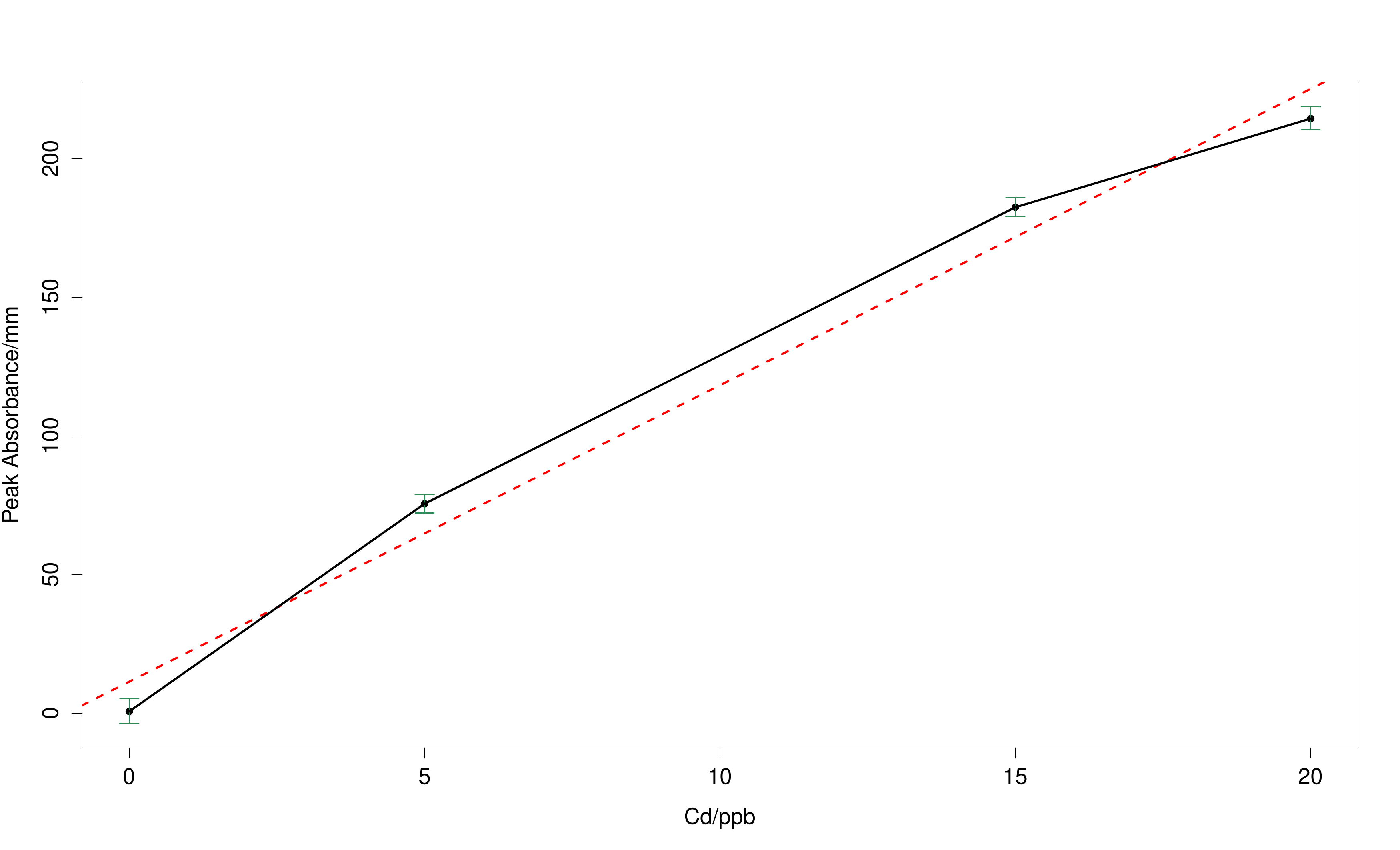}
\caption{Linear (dotted) and quadratic calibration curves for Cd using four standard references. Indicated are 95\% credible intervals of the estimated absorbances.}
\label{fig:Calib_Ex}
\end{center}
\end{figure}
In Lundberg and de Mar\'{e}'s (1980) example graphite furnace atomic absorption spectroscopy (GFAAS) was used to determine trace amounts of cadmium (Cd) in water samples. In the first stage of experimentation two $\mu$l volumes of standard solutions containing $0, 5, 10, 15$, and $20$ ppb (parts per billion) of Cd were injected into a graphite furnace (Varian Techtron AA-6 spectrophotometer supplied with a Carbon Rod Atomizer model 63).
\begin{table}[!ht]
\centering
\captionof{table}[]{Cd signals obtained when atomizing standards
}
\begin{tabular}{l|rrrrrr}
\toprule
\multicolumn{1}{l}{{{Concentration/ppb}}}\vline
& \multicolumn{6}{c}{{{Peak Absorbance/mm}}}\\
  \midrule
0    & 0    & 1     & 1     & 0     & 1      &\\ 
5    & 74  & 74   & 78   & 78   & 76    & \\ 
15  &183 & 184 & 178 & 183 & 184  & \\ 
20  & 217 & 215 & 213 & 218 & 210 & 215  \\
   \hline
\end{tabular}\label{tab:cal_ex}
\end{table}
The transient absorbance signals obtained when atomizing the standards were recorded with a stripchart recorder. Each standard was run several times and a plot of peak absorbance (in mm) vs. concentration was made using the data in Table \ref{tab:cal_ex} (See Figure \ref{fig:Calib_Ex}). At the second stage of experimentation the fifth reference (10 ppb) was used as an unknown sample to test the validity of the method with peak absorbance measurements of 135,142,132, 141, and 136. Lundberg and de Mar\'{e} (1980) report an approximate 95\% confidence interval for the unknown concentration with 10 being the true value as $\left[9.7, 10.3\right]$.\\
\begin{figure}[htbp]
\begin{center}
\includegraphics[width = 14cm, height = 8cm]{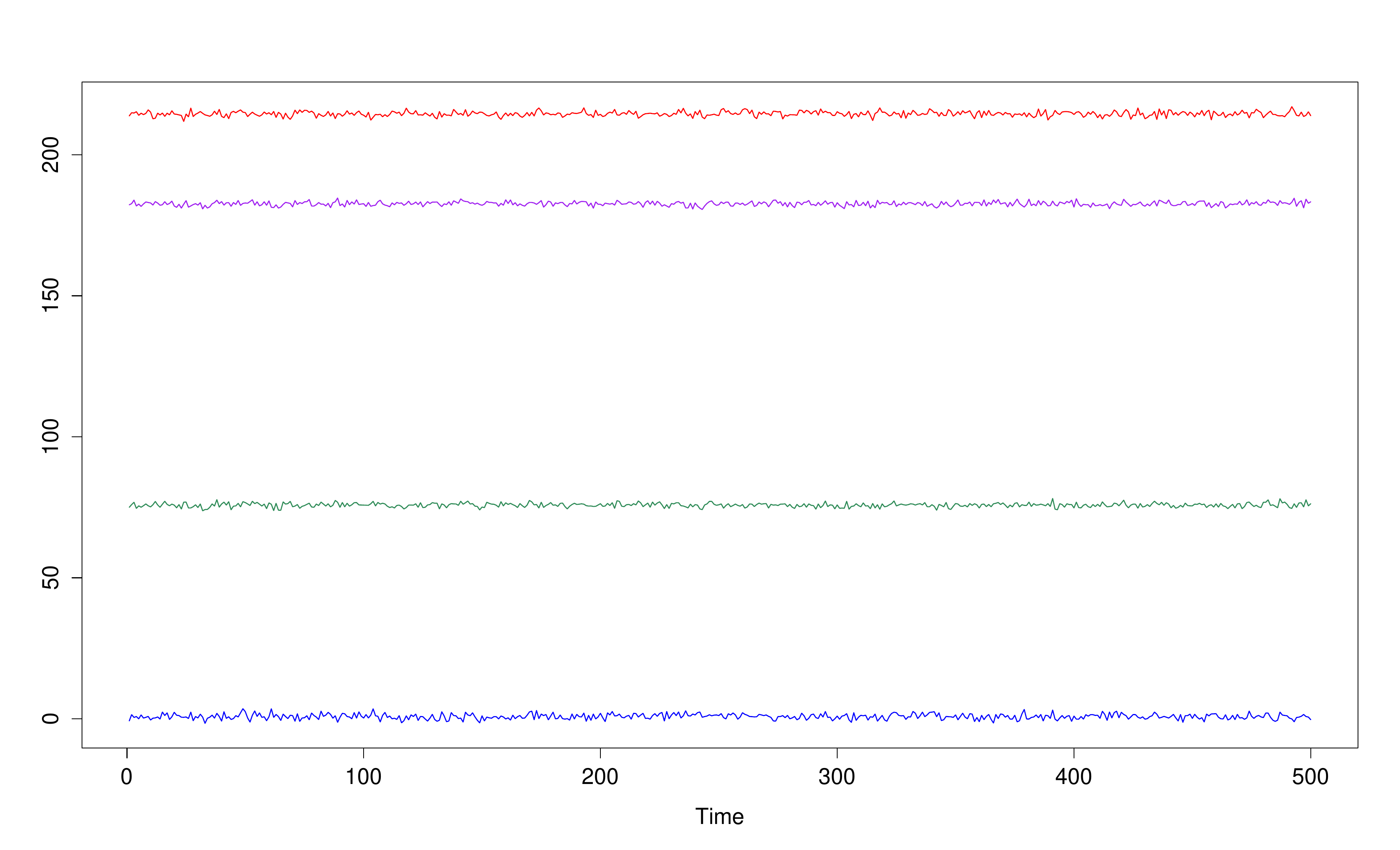}
\caption{Peak Absorbance measurements: Concentration 0ppb (blue); Concentration 5ppb (green); Concentration 15ppb (violet); and Concentration 20ppb (red).}
\label{fig:Calib_Ex2}
\end{center}
\end{figure}
\indent We extend Lundberg and de Mar\'{e}'s (1980) example by generating 500 simulated peak absorbance measurements for the four standard solutions of Cd (see Figure \ref{fig:Calib_Ex2}). The repeated measures for the peak absorbance were generated given a multivariate normal distribution ${\boldsymbol \beta}_{t}$ with mean vector
\[ 
{\boldsymbol \mu} =
\left[ \begin{array}{c}
0.72 \\
16.448 \\
-0.288 \end{array} \right]\]
and variance-covariance matrix
 \[ 
{\boldsymbol \Sigma} = \sigma^{2}
\left[ \begin{array}{ccc}
0.17966  & -0.03435 & 0.00131 \\\ 
-0.03435 & 0.01473  & -0.00069 \\ 
0.00131  & -0.00069 & 0.00003 \\  \end{array} \right],\]
 where $\sigma^{2} = 4.7$, the residual variance from the ordinary least squares fit of the original data.
\begin{figure}[!htb]
\begin{center}
\includegraphics[width=14cm,height=8cm]{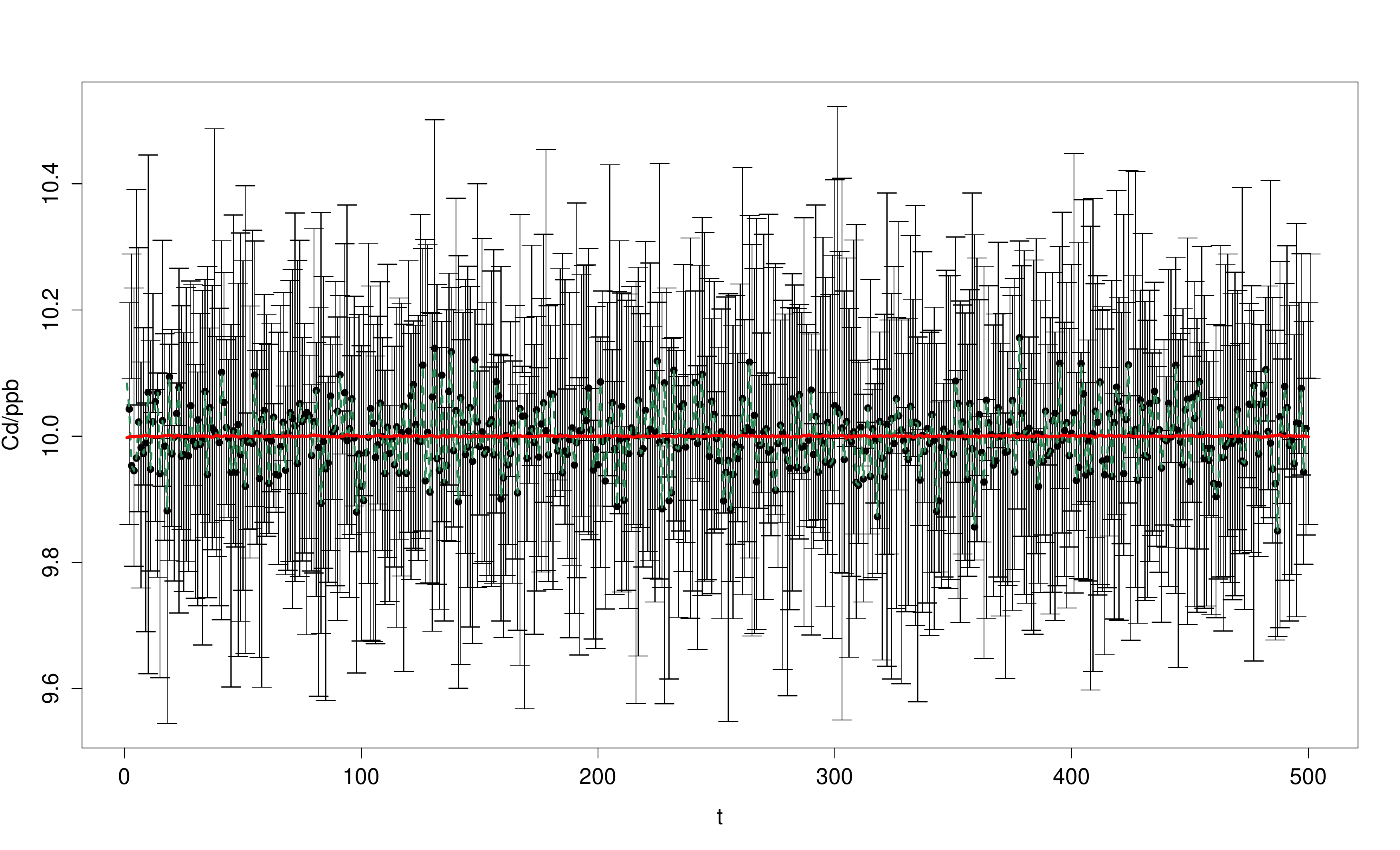}
\caption{Time series of Cd estimates (black dots connected by green lines) with corresponding 95\% credible interval at time $t$. The "true" Cd of 10 ppb is given by the red line.}
\label{fig:Calib_Sol}
\end{center}
\end{figure}
%
%
The time series of the posterior estimates for Cd is given in Figure \ref{fig:Calib_Sol}. The square root of the mean squared error is 0.0025 with an average 95\% credible interval [9.8~10.2] which is shorter than the 95\% confidence interval of [9.7~10.3] reported by Lundberg and de Mar\'{e} (1980).\\

	\subsection*{Example 2}

\begin{figure}[!htb]
\begin{center}
\includegraphics[width=14cm,height=8cm]{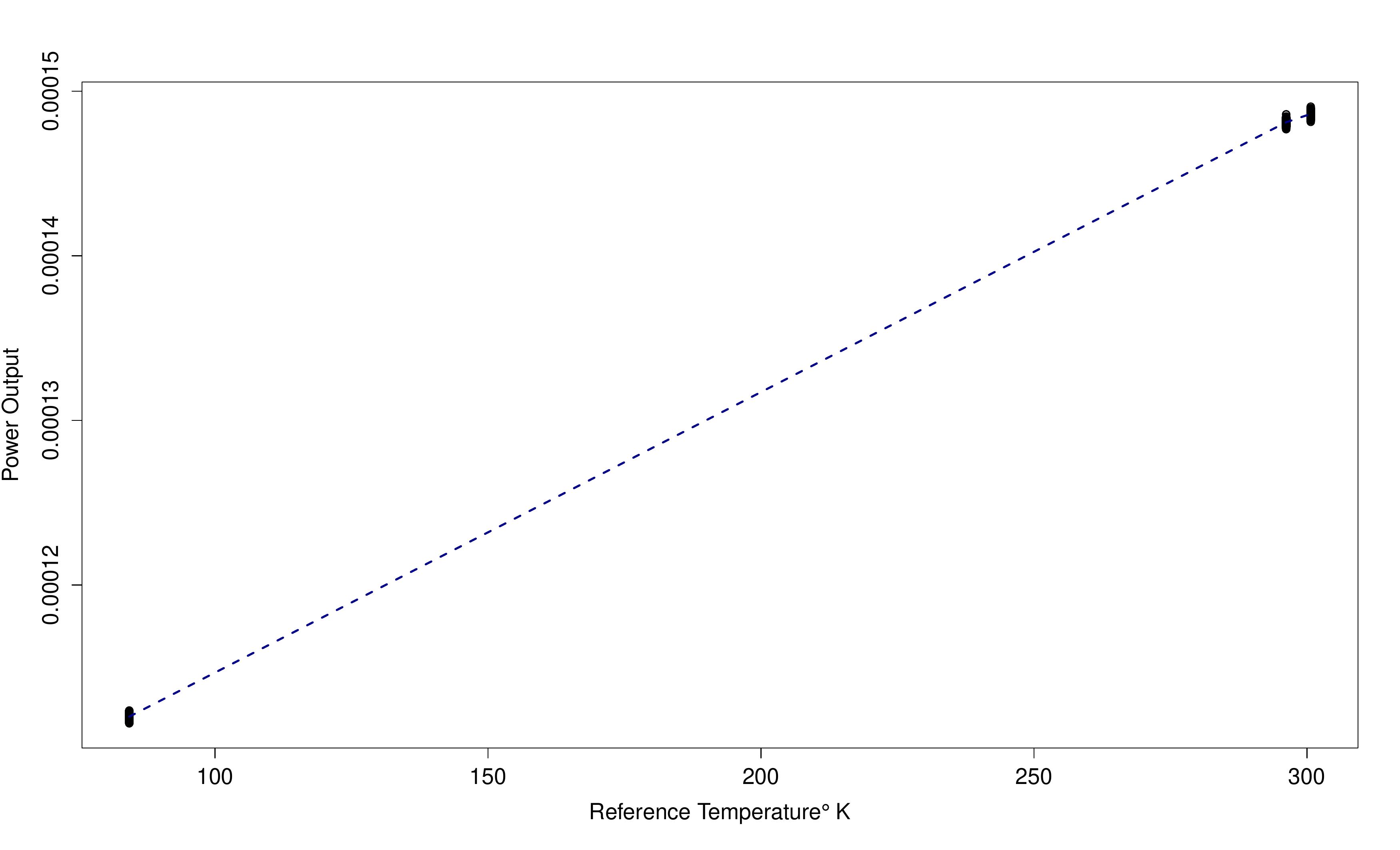}
\caption{3-Point Calibration model used to detect nonlinearity. Nonlinear calibration curve (black dots connected by dotted red line) goes through three reference measurements collected over time.}
\label{fig:NIST1}
\end{center}
\end{figure}

\begin{figure}[!htb]
\begin{center}
\includegraphics[width=14cm,height=8cm]{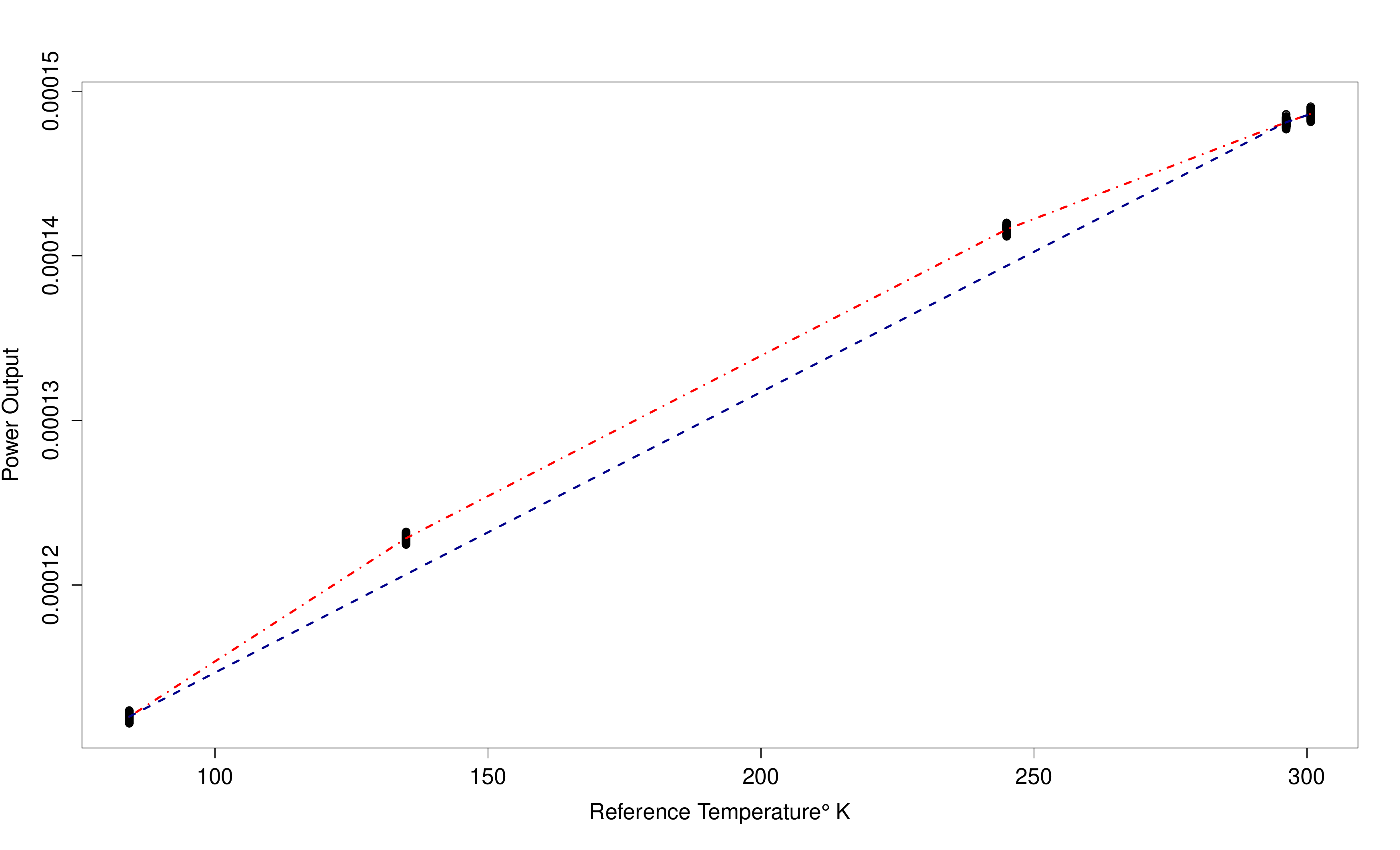}
\caption{5-Point Calibration model versus 3-Point Calibration model. The calibration curve for the 3-point model is shown by the dotted line and the calibration curve for the 5-point model is shown by the dash-dotted line.}
\label{fig:NIST2}
\end{center}
\end{figure}

\indent We turn to an example in microwave radiometry to also demonstrate the usefulness of the proposed dynamic calibration approach. Engineers and scientist commonly use microwave radiometers to measure electromagnetic radiation. This radiant power is emitted by some source or a particular surface such as ice or land surface. Radiometers are very sensitive instruments that are capable of measuring extremely low levels of radiation. The transmission source of the radiant power is the target of the radiometers antenna. When a scene, such as terrain, is observed by a microwave radiometer, the radiation received by the antenna is partly due to self-emission by the scene and partly due to the reflected radiation originating from the surroundings (Ulaby et al. 1981). This source may be cosmic background radiation, ocean surface, or a heated surface used for the purpose of calibration.\\
\indent Calibration is required due to the fact that the current electronic hardware is unable to maintain a stable input/output relationship. For space observing instruments, stable calibration without any drifts is a key to detect proper trends of climate (Imaoka {\it et al.} 2010). Due to problems such as amplifier gain instability and exterior temperature variations of critical components that may cause this relationship to drift over time (Bremer 1979). During the calibration process, the radiometer receiver measures the voltage output power $v(t)$, and its corresponding input temperature of a known reference. Two or more known reference temperatures are needed for calibration of a radiometer. Ulaby {\it et al.} (1981); Racette and Lang (2005) state that the relationship between the output, $v(t)$ and the input, $T_{A}$ is approximately linear, and can be expressed as	
\begin{equation*}
\hat{T}_{A} = \hat{\beta}_{0} + \hat{\beta}_{1}v(t) \label{eqn:calibequation} 
\end{equation*}
where, $\hat{\beta}_{0}$ and $\hat{\beta}_{1}$ are the least square estimates for the regression paramters, $\hat{T}_{A}$ is the estimated value of the brightness temperature and $v(t)$ is the observed output voltage. Using this relationship, the output value, $v(t)$, is used to derive an estimate for the input, $T_{A}$ (Racette and Lang, 2005).\\
\indent It is of interest to develop a calibration approach that can detect gain abnormalities, and/or correct for slow drifts that affect the quality of the instrument measurements. To demonstrate the dynamic approach in terms of application appeal, Rivers and Boone (2014) used the dynamic approach to characterize a calibration target over time for a microwave radiometer. The data used for this example was collected during a calibration experiment that was conducted on the Millimeter-wave Imaging Radiometer (MIR) (Racette et al. 1995). The purpose of the experiment was to validate predictions of radiometer calibration.\\
\begin{figure}[!htb]
\begin{center}
\includegraphics[width=14cm,height=8cm]{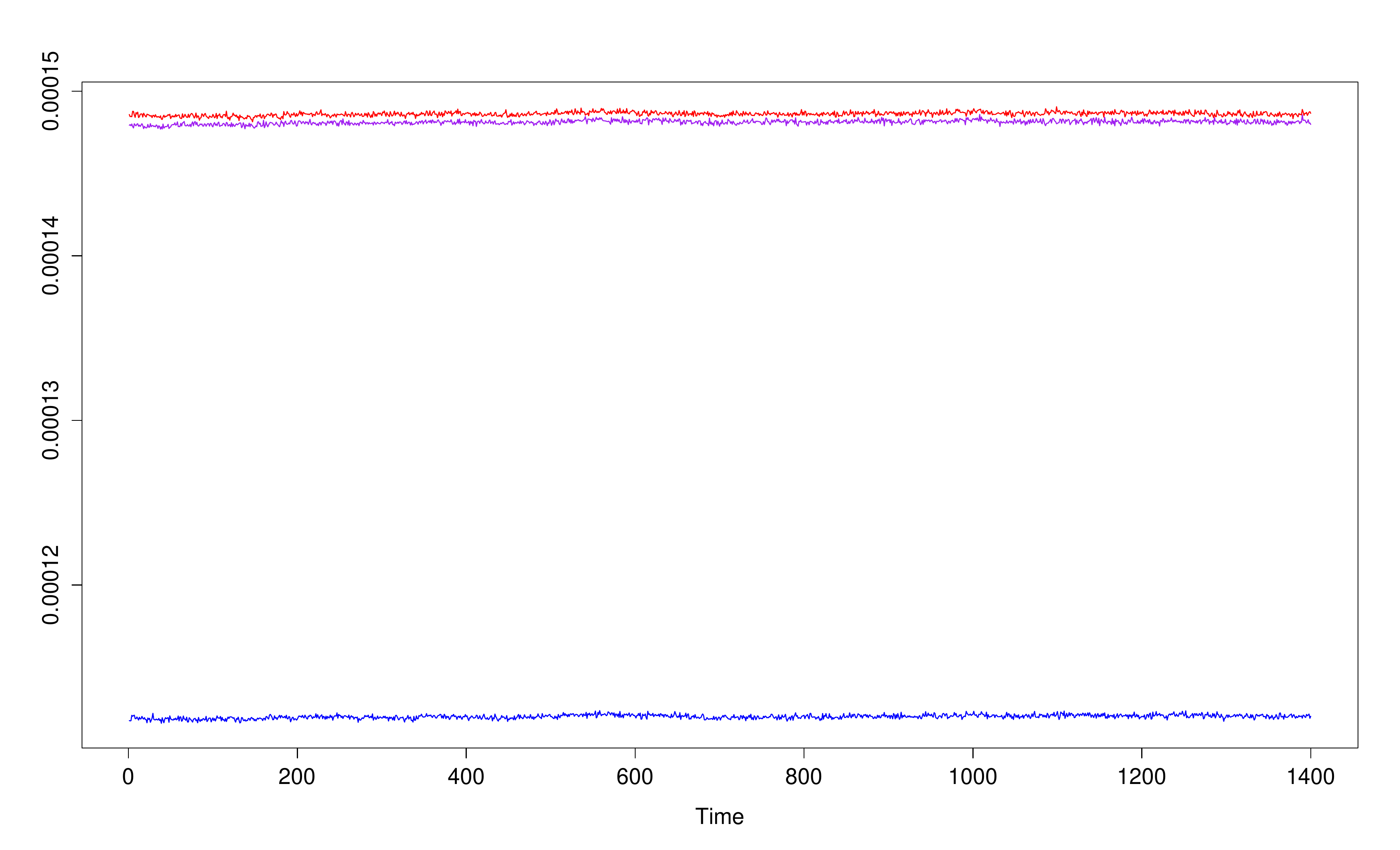}
\caption{Power output measurements: V$_{cryo}$ (blue); V$_{amb}$ (violet); and V$_{warm}$ (red)}
\label{fig:NIST_data1}
\end{center}
\end{figure}
\indent We extend the work of Rivers and Boone (2014) to the nonlinear model by examining a data set similar to one that one be created by a laboratory bench calibration experiment. The National Institute of Standards and Technology (N.I.S.T.) conducted a bench calibration experiment designed at studying calibration methods, due to intellectual property rights the data is not free to use, so with the aid of computers we simulated data with the attributes of the N.I.S.T. data. The data set consist of three temperature references and the corresponding three power output measurements collected over 1400 time periods. The references temperatures are as follows:
\begin{itemize}
	\item T$_{cryo}   = 84.3^{\circ} K$
	\item T$_{amb}   = 296.2^{\circ} K$
	\item T$_{warm} = 300.7^{\circ} K$
\end{itemize}
with summary statistics for the corresponding output measurements as
\begin{table}[ht]
\centering
\begin{tabular}{rccc}
  \hline
 & V$_{cryo}$ & V$_{amb}$ & V$_{warm}$ \\ 
  \hline
$\bar{v}$               & 0.0001120096 & 0.0001481137 & 0.0001486190 \\ 
$\sigma_{v}$ & 0.0000001280 & 0.0000001308 & 0.0000001236 \\ 
   \hline
\end{tabular}
\end{table}\\
The time series plots for the observed output measurements are provided in Figure \ref{fig:NIST_data1}.\\
\indent In Figure \ref{fig:NIST1} we plot the output measurements against the reference temperatures and show the calibration curve by the dotted line. One may suggest that a linear fit is appropriate but the experiment that inspired this data was created such that the quadratic term $\beta_{2t}$ would be significant.\\
\indent We demonstrate the dynamic calibration method on the simulated radiometer data by setting $y_{0t} = 0.0001347169$ which corresponds to $x_{0t} = 200^{\circ} K$ given a stable system without any drift in the dynamic regession parameters. An assessment of the dynamic method is conducted at the end of the time series when $t=1400$. After employing Algorithm \ref{algo:DynCal} of the data we get a mean value for $\hat{x}_{0t}$ across time as 199.5695 with a standard deviation $\sigma_{x_{0t}}=1.53135$. It is of interest to now if the addition of references would improve the estimation process because the placement of the references in the experiment conducted at NIST may not completely capture the degree of the nonlinearity. In Figure \ref{fig:NIST2} we added two additional reference measurements. We place the additional measurements as follow:
\begin{itemize}
	\item T$_{135}   = 135^{\circ} K$
	\item T$_{245}   = 245^{\circ} K$
\end{itemize}
with the corresponding output measures listed in the table below
\begin{table}[ht]
\centering
\begin{tabular}{rcc}
  \hline
 & V$_{135}$ & V$_{245}$ \\ 
  \hline
$\bar{v}$               & 0.0001228344 & 0.0001415994  \\ 
$\sigma_{v}$ & 0.0000001257 & 0.0000001233  \\ 
   \hline
\end{tabular}
\end{table}\\
The dynamic calibration method was conducted on the data set with the inclusion of the additional reference measurements yielding a mean value across time of $\hat{x}_{0t}=200.1122$ which is closer to the assumed true temperature measure of $200^{\circ} K$ and a standard deviation of $\sigma_{x_{0t}}=0.4042387$. In Table \ref{tab:Rad_Sim} we compare the performance criterions from Section \ref{sec:Sim_Study} side-by-side for the data when using a 3-point calibration model and 5-point calibration model.\\
\begin{table}[!bht]
\centering
\captionof{table}[Comparison of 3-point calibration model versus the 5-point calibration model]{
Summary of the 3-point calibration model versus the 5-point calibration model. 
}
\begin{tabular}{rcccccc}
\toprule
Model & MSE & IW & CP \\ 
  \midrule
3-Point   & 2.5130986 & 2.2480498 & 0.4450423 \\ 
5-Point   & 0.1602064 & 1.8286866 & 0.9699140 \\ 
   \hline
\end{tabular}\label{tab:Rad_Sim}
\end{table}
\indent In Table \ref{tab:Rad_Sim} it is easy to see that the inclusion of the two reference measurements greatly improved the estimation of the calibration distributions across time. This is evident by the decrease in the mean square error. The interval width is shorter for the 5-point model because as stated before that standard deviation when the 2 measurements were added is significantly smaller than when not including them. The greatest improvement took place in the coverage probability ($CP$) because it nearly doubled by pacing the references between the endpoint measurements.\\
\indent The results shown in this section indicate that, dynamic calibration can be used to determine concentrations, reference temperatures, or any other unknown measurement source where time period of calibration is significant. Until Rivers and Boone (2014) introduced the dynamic method statistical calibration has been considered from a static point of view. To add to the discussion of the dynamic Bayesian nonlinear calibration, next we consider a case where concern may arise when the observed measurement $y_{0t}$ approaches the vertex. This sort of issue is considered to be problematic, thus we are concerned with how the dynamic method performs in this scenario. 

\section{Future works and other considerations}\label{sec:future}

\begin{figure}[!htb]
\begin{center}
\includegraphics[width=14cm,height=8cm]{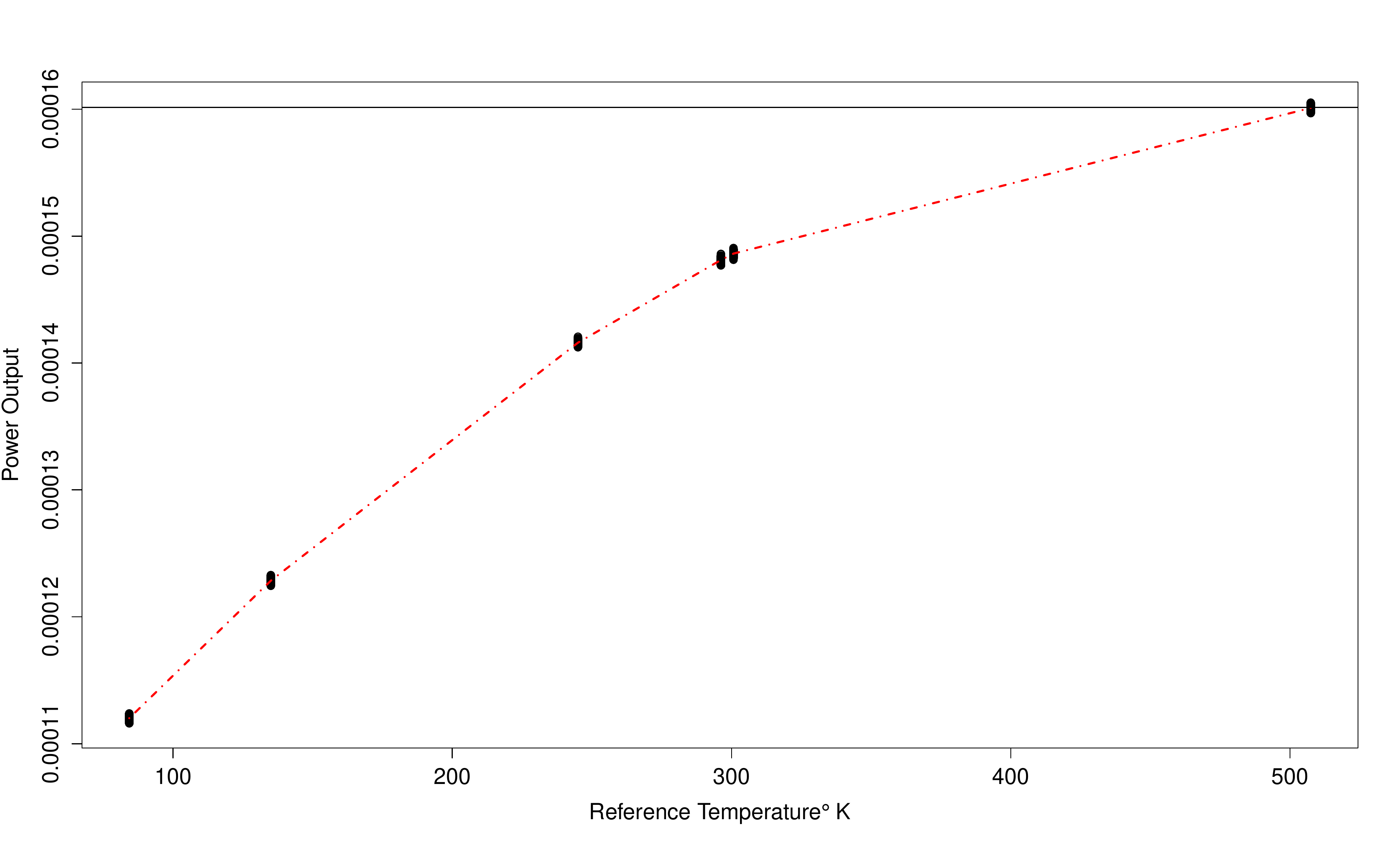}
\caption{Quadratic calibration model (dash-dotted line) with observation $y_{0t}$ near vertex. }
\label{fig:vertex1}
\end{center}
\end{figure}

\begin{figure}[!htb]
\begin{center}
\includegraphics[width=14cm,height=8cm]{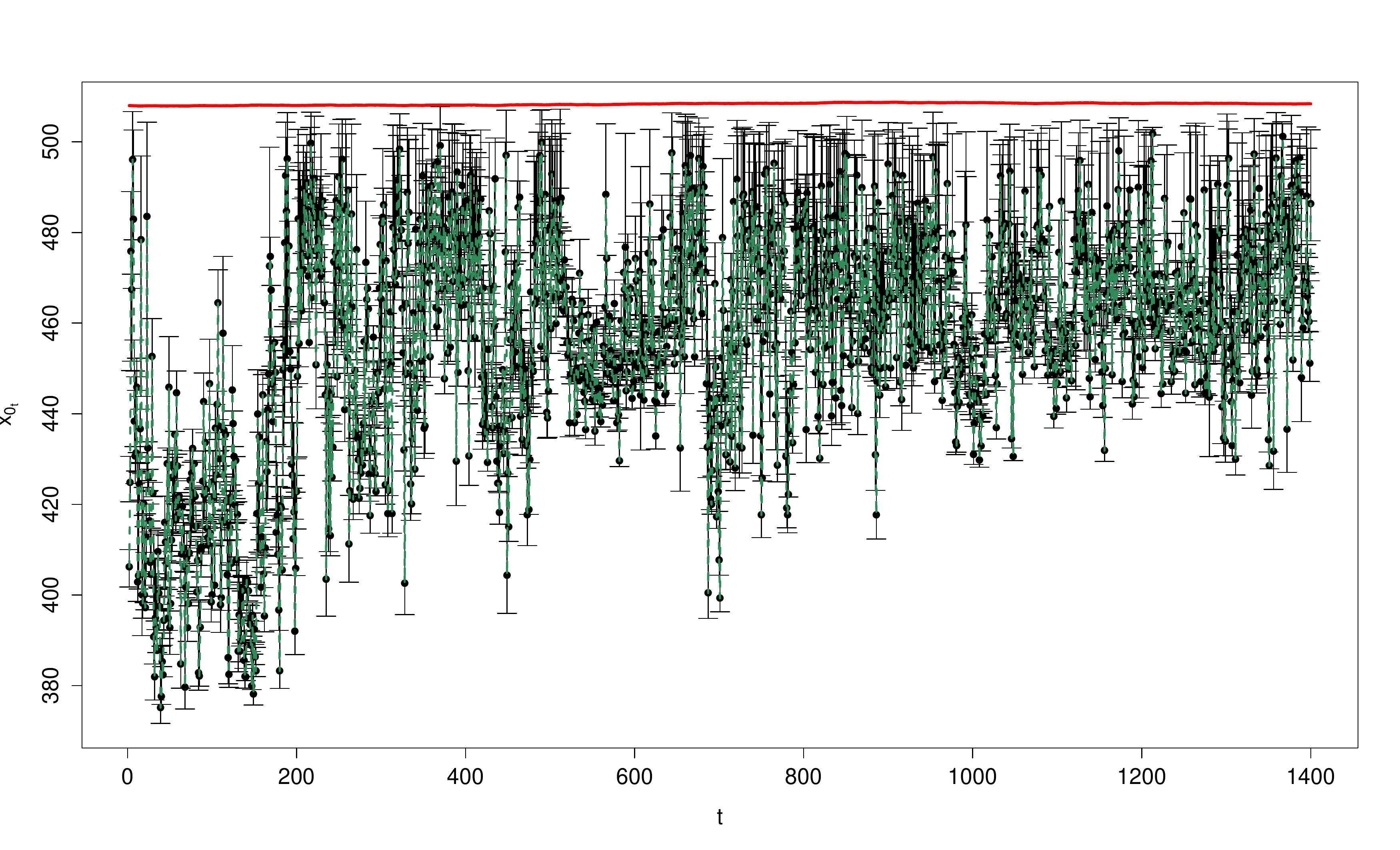}
\caption{Quadratic calibration model (dash-dotted line) with observation $y_{0t}$ near vertex. }
\label{fig:vertex2}
\end{center}
\end{figure}

\begin{figure}
\centering
\begin{subfigure}{0.5\textwidth}
  \centering
  \includegraphics[width=1.0\linewidth]{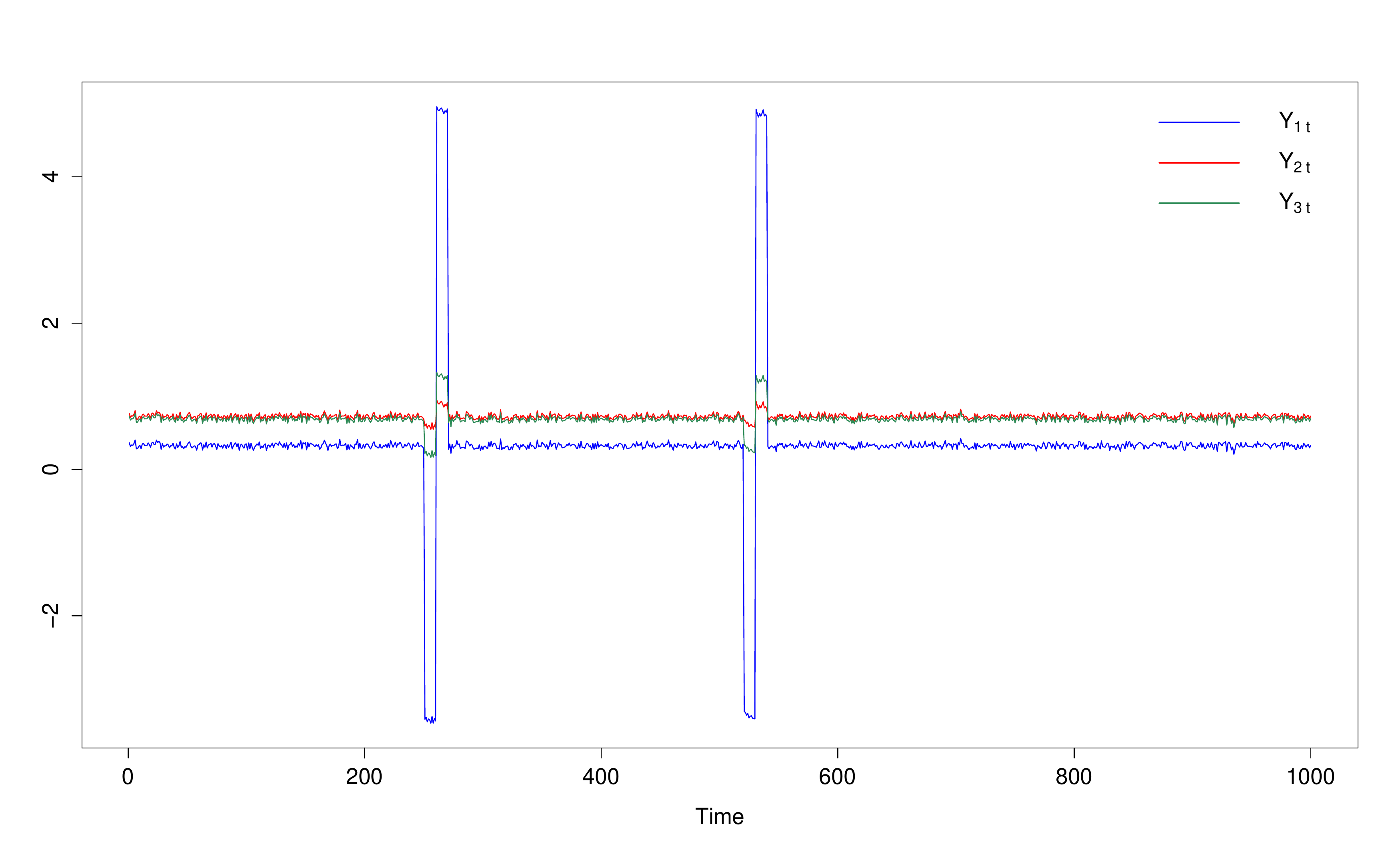}
  \caption{Time series of 3 observation measurements with shocks at $t = 250$ and $t = 520$.}
  \label{fig:shock1}
\end{subfigure}%
~
\begin{subfigure}{0.5\textwidth}
  \centering
  \includegraphics[width=1.0\linewidth]{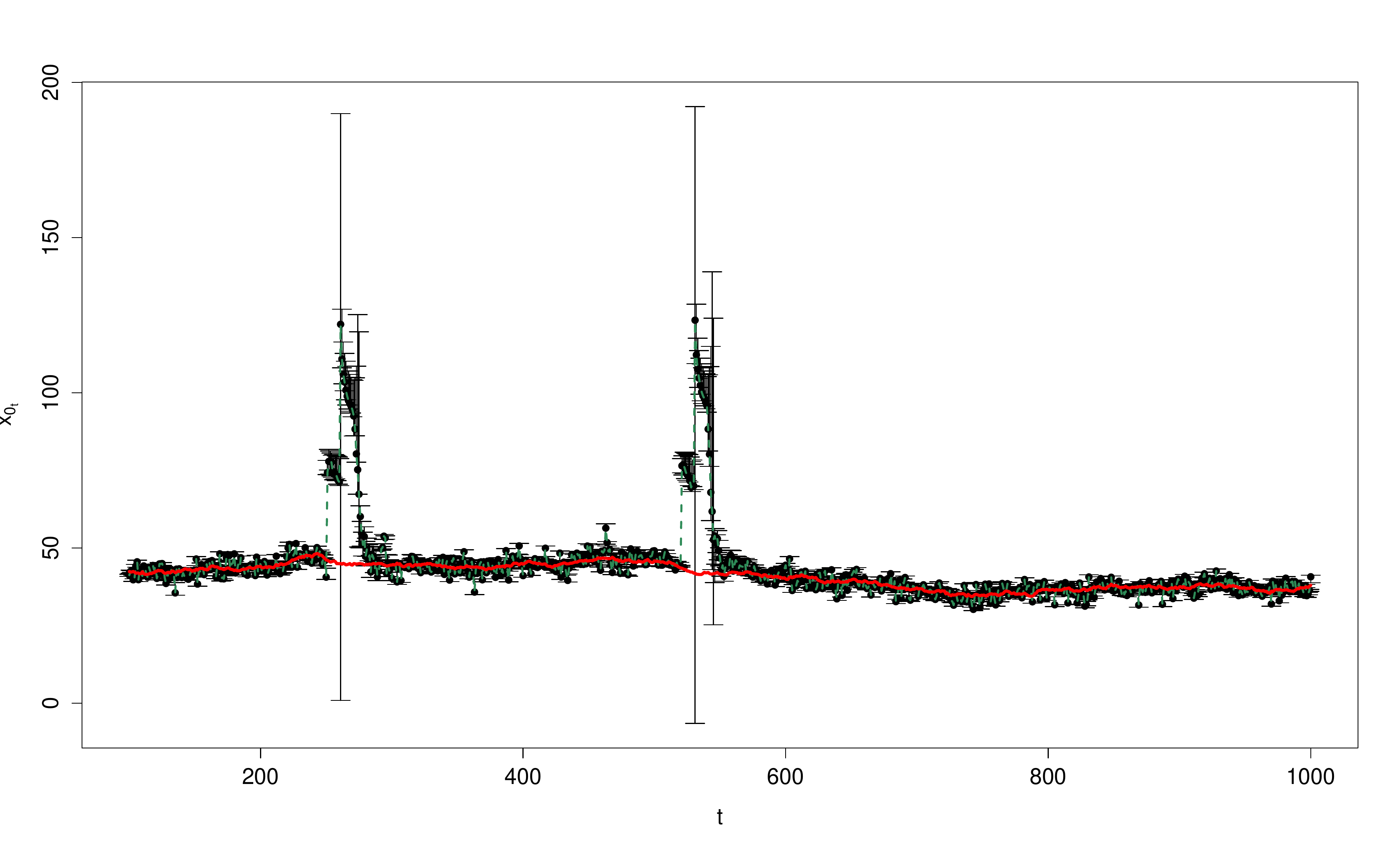}
  \caption{Calibration distributions in the presence of system disturbances.}
  \label{fig:shock2}
\end{subfigure}
\caption{System disturbance for 20 time periods.}
\label{fig:shock_ex1}
\end{figure}

\indent Before concluding this work we would be remiss not to consider the case when the observed measurement $y_{0t}$ approaches the vertex. We would like to understand how the calibration method will perform under such a condition. To understand the behavior of the vertex in a dynamic sense we look at the time-varying quadratic equation in vertex form,
\begin{equation}
y_{t} = \hat{a}_{t}(x - \hat{h}_{t})^{2} + \hat{k}_{t}
\end{equation}
where
\begin{eqnarray*}
\hat{a}_{t} &=& \hat{\beta}_{2t},\\
\hat{h}_{t} &=& \frac{-\hat{\beta}_{1t}}{2\hat{\beta}_{2t}},\\
\hat{k}_{t} &=& \hat{\beta}_{0t} - \frac{-\hat{\beta}^{2}_{1t}}{4\hat{\beta}_{2t}},  
\end{eqnarray*}
and the time-varying vertex is $(\hat{h}_{t},~\hat{k}_{t})$.\\
\begin{figure}[!htb]
\begin{center}
\includegraphics[width=14cm,height=8cm]{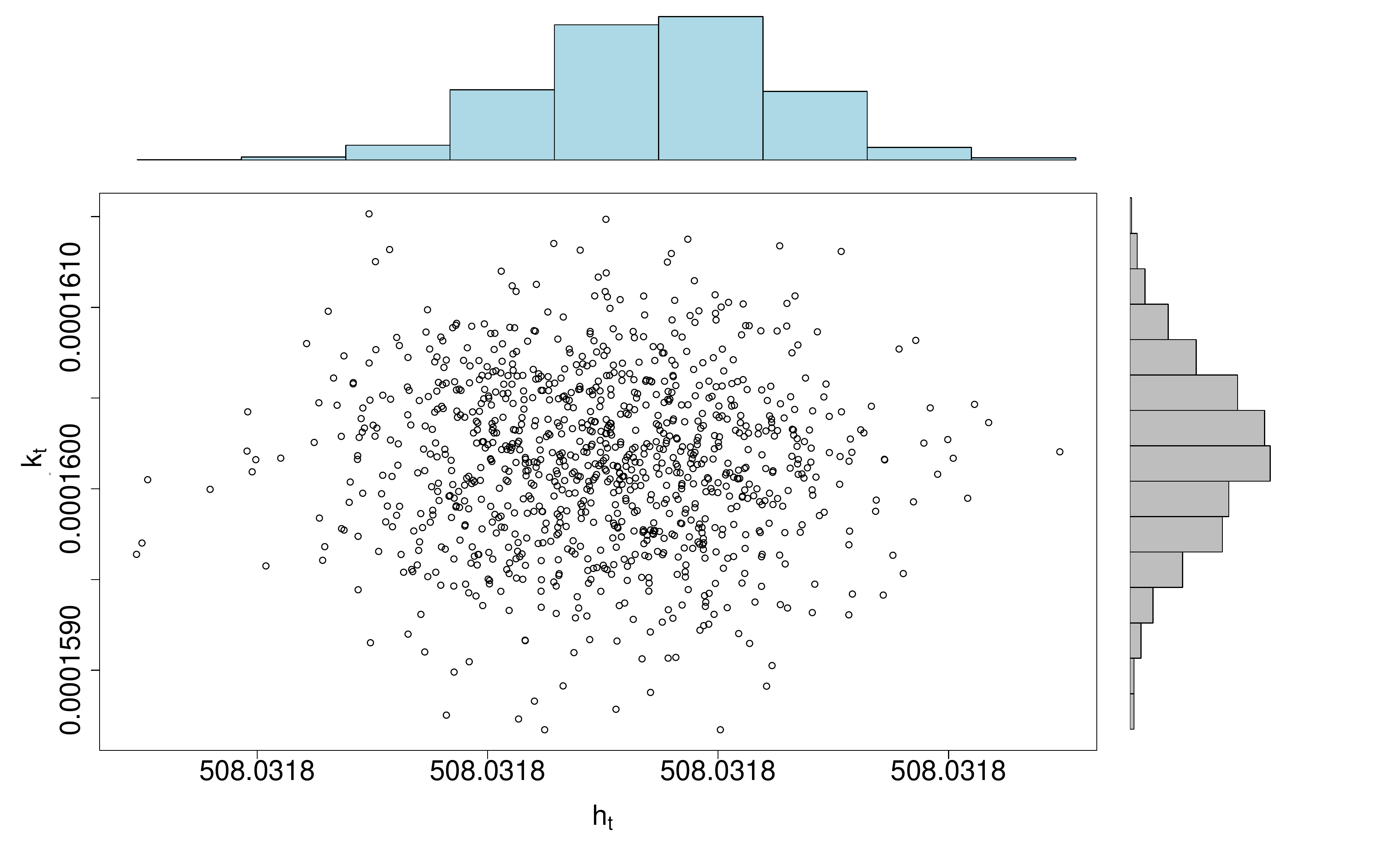}
\caption{Joint distribution of $(\hat{h}_{t},~\hat{k}_{t})$ with marginal distributions of $\hat{h}_{t}$ and $\hat{k}_{t}$ }
\label{fig:vertex}
\end{center}
\end{figure}
\indent The estimated dynamic regression parameters are used to derive the joint distribution of $(\hat{h}_{t},~\hat{k}_{t})$ given in Figure \ref{fig:vertex}. So, we extend the simulated radiometer data previously used in Section \ref{sec:application} by placing a reference measurement at $508^{\circ} K$. A time series of observed power outputs $y_{0t}$ is generated from the following quadratic equation:
\begin{equation}
y_{0t} = \hat{\beta}_{0t} + \hat{\beta}_{1t}(508) + \hat{\beta}_{2t}(508)^{2},\label{eq:section5}
\end{equation}
where $\hat{\beta}_{0t}$, $\hat{\beta}_{1t}$ and $\hat{\beta}_{2t}$ are the estimates of the time dependent regression parameters. In Figure \ref{fig:vertex1} the black horizontal line represents the possible maximum power output $V_{t}$ given the vertex is $(\hat{h}_{t},~\hat{k}_{t})$ at time $t$.\\
%
%
%
\indent We used dynamic calibration method described by Algorithm \ref{algo:DynCal} to examine just how the method would perform at estimating the temperature value of $508^{\circ}K$. In Figure \ref{fig:vertex2} we see that the method performs poorly from a mean squared error and coverage probability point of view. The method is restricted from deriving distributions that violate the quadratic behavior by not calculating credible intervals that go beyond any real possible value, meaning that  the upper credible limit $x_{0t}^{U}$ will be less than the true value for $x_{0t}$ with $P( x_{0t}^{U} <  x_{0t})=1$.
Clearly, it is not advisable to perform a statistic calibration experiment so close to the vertex. Calibration, whether static or dynamic is best conducted when the calibration target is near the center of the references domain space.\\ 
\indent Finally, we consider an extreme case where some sort of shock is imposed with the system of the calibration device that causes drastic shifts in the observation data at random points in time. It is of interest to understand how the dynamic approach handles such a case. As before, we model the dynamic system in vertex form,
\begin{equation}
y_{t} = \hat{s}_{t}(x - \hat{h}_{t})^{2} + \hat{k}_{t}
\end{equation}
where
\begin{eqnarray*}
\hat{s}_{t} &=& \hat{\beta}_{2t}\gamma_{t},\\
\hat{h}_{t} &=& \frac{-\hat{\beta}_{1t}}{2\hat{\beta}_{2t}},\\
\hat{k}_{t} &=& \hat{\beta}_{0t} - \frac{-\hat{\beta}^{2}_{1t}}{4\hat{\beta}_{2t}},  
\end{eqnarray*}
and the random multiplicative disturbance is denoted as $\gamma_{t}$. In Figure \ref{fig:shock1} we have three time series measurements, denoted as $Y_{1t}$, $Y_{2t}$, and $Y_{3t}$, from a calibration experiment that is taken from the reference temperature measurements of $20^{\circ}K, 90^{\circ}K$, and $100^{\circ}K$. We see, that at times $t = 250$ and $t = 520$ a random shock has been imposed on the system thus causing mean-shifting glitches in the observation data. Each of the random shock periods last for exactly 20 time periods in Figure \ref{fig:shock1}.
\begin{figure}
\centering
\begin{subfigure}{.5\textwidth}
  \centering
  \includegraphics[width=1.0\linewidth]{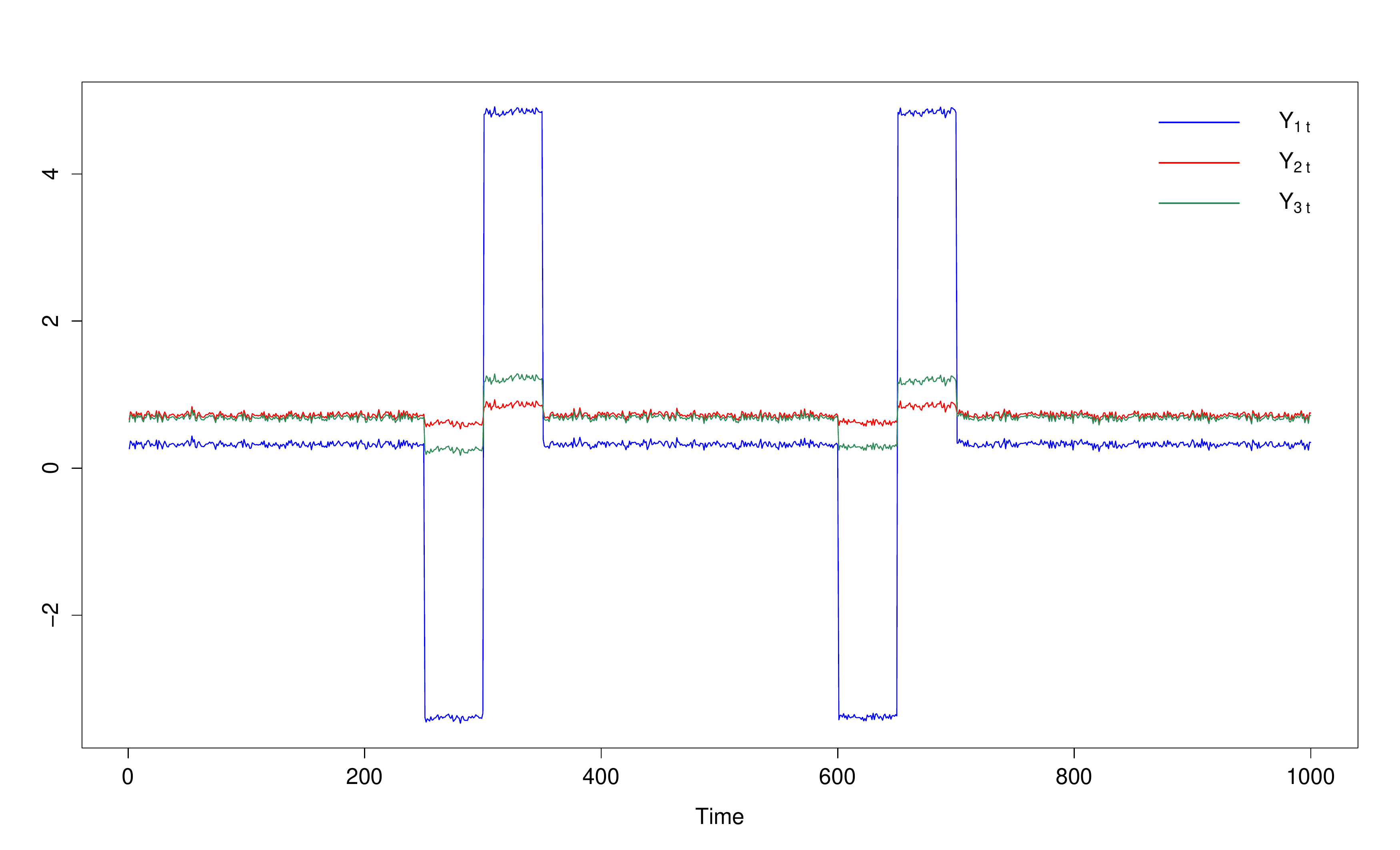}
  \caption{Time series of 3 observation measurements with shocks at $t = 250$ and $t = 520$.}
  \label{fig:shock1b}
\end{subfigure}%
~
\begin{subfigure}{.5\textwidth}
  \centering
  \includegraphics[width=1.0\linewidth]{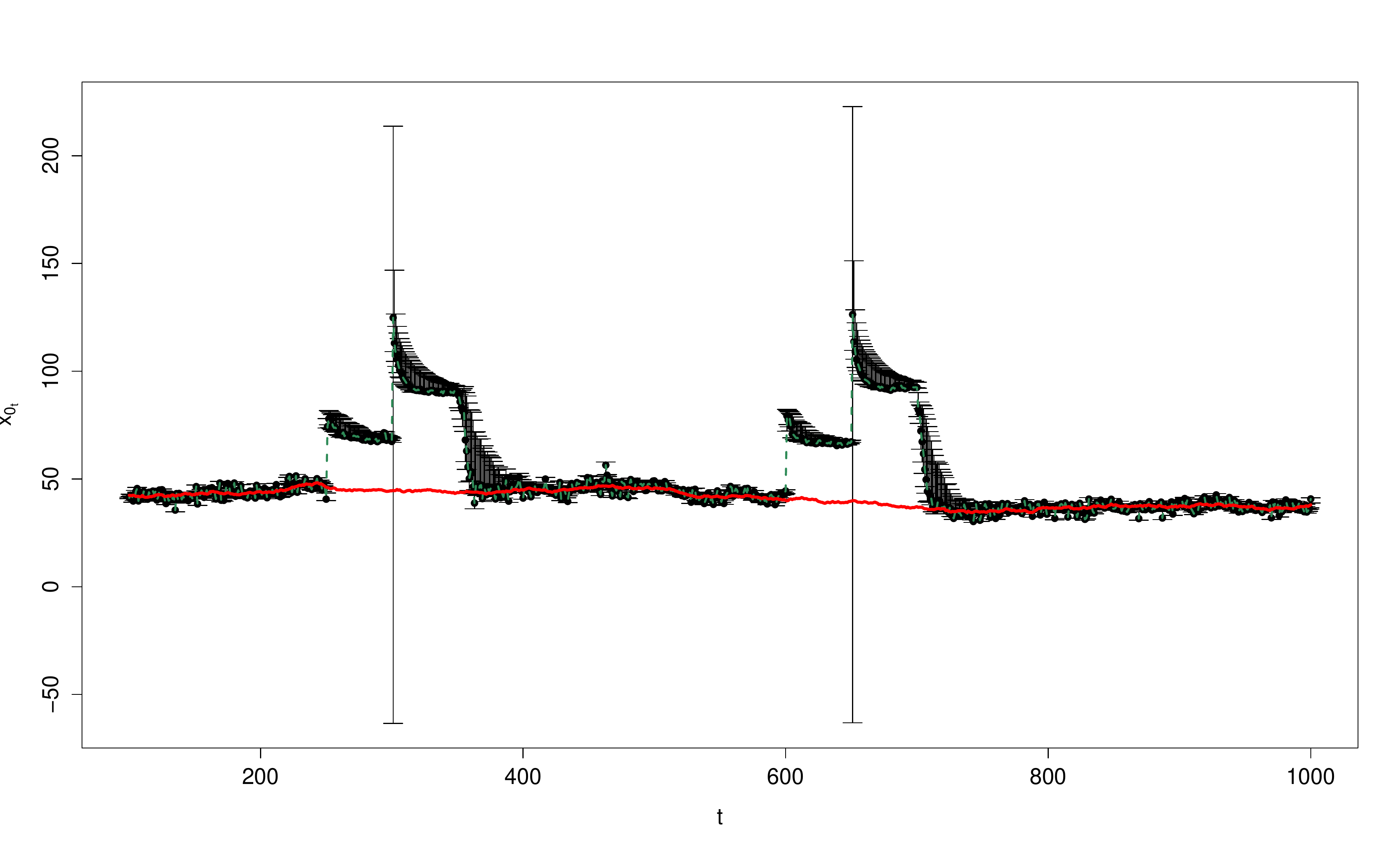}
  \caption{Calibration distributions in the presence of system disturbances.}
  \label{fig:shock2b}
\end{subfigure}\caption{System disturbance for 100 time periods.}
\label{fig:shock_ex2}
\end{figure}
 The disturbance to the system get incrementally greater as we move away from the vertex, one would see larger fluctuations for the observations for the lower reference measurement than for the higher reference measurements. This supports the conclusions made by Fran\c{c}ois et al. (2004). We see in Figure \ref{fig:shock2} that the calibration distributions credible intervals grow wider during the interrupted period as the dynamic method attempts to learn the behavior of the regression relationship. A consequent of the glitches is that the coverage probability will decrease and the interval widths will increase.\\
\indent To get a better understand of the behavior of the method we extend the time frame of the system disturbance from 20 time periods to 100 time periods where half of the time the disturbance is negative and the other half it is positive. In Figures \ref{fig:shock1b} and \ref{fig:shock2b} we see that the disturbance in the system is translated in dynamic method. Eventhough the dynamic methods coverage probability is drastically lessened by the glitch it still outperforms the static method as the static quadratic calibration method is unable to derive estimates across time if there is instability in the parameters. Future work is to investigate the dynamic calibration methods from a multiple multivariate calibration point of view. In such settings we not only establish the inter-device calibration relationship to perform calibration over time of the particular instrument but we establish a dependency among the instruments as well. 

%
%





\end{document}